\documentclass[a4paper,fleqn]{cas-sc}

\usepackage{subfigure}  
\usepackage{graphicx}
\usepackage{booktabs}
\usepackage{array}
\usepackage{enumitem}
\usepackage{float}
\usepackage[export]{adjustbox} 
\usepackage{placeins} 

\AtBeginDocument{%
  \setcounter{topnumber}{3}%
  \setcounter{bottomnumber}{2}%
  \setcounter{totalnumber}{5}%
}

\usepackage{amsthm}
\usepackage[numbers]{natbib}

\usepackage{algorithm}
\usepackage{algorithmic}	
\usepackage[english]{babel}

\newcommand{\ReP}{\mathcal{R}\mathrm{e}}
\newcommand{\ImP}{\mathcal{I}\mathrm{m}}

\def\tsc#1{\csdef{#1}{\textsc{\lowercase{#1}}\xspace}}
\tsc{WGM}
\tsc{QE}
\tsc{EP}
\tsc{PMS}
\tsc{BEC}
\tsc{DE}

\ExplSyntaxOn
\makeatletter
\cs_gset:Npn \__first_footerline: { \group_begin: \small \sffamily \__short_authors: \group_end: }
\cs_gset:Npn \__first_foot: { \parbox[t]{\textwidth}{ \rule{\textwidth}{.2pt}\\ \__first_footerline: } }
\cs_gset:Npn \__cas_foot: { \parbox[t]{\textwidth}{ \rule{\textwidth}{.2pt}\\ \__first_footerline: } }
\ps@cas
\makeatother
\ExplSyntaxOff

\ExplSyntaxOn
\makeatletter
\NewDocumentEnvironment{inlinetable}{}
  {\par\medskip\noindent\begin{minipage}{\linewidth}
   \cs_set_eq:NN \@makecaption \__make_tbl_caption:nn
   \def\@captype{table}\centering\sffamily\small}
  {\end{minipage}\par\medskip}
\makeatother
\ExplSyntaxOff

\begin{document}
	\let\WriteBookmarks\relax
	\def\floatpagepagefraction{1}
	\def\textpagefraction{.001}
	\shorttitle{FCV-PICRN for Gappy and Noisy Flow-Field Reconstruction}
	\shortauthors{Barman, Ray}
	
	\title [mode = title]{Fourier-Feature-Embedded Complex-Valued Physics-Informed Convolutional Residual Networks for Sparse and Gappy Noisy Flow-Field Reconstruction}                      
	\author[1]{Biswanath Barman}
    \ead{biswa.nit.rourkela@gmail.com}
        
        \author[1]{Rajendra K. Ray$^{*}$} 
        \cortext[cor1]{Corresponding author}
        \ead{rajendra@iitmandi.ac.in}
        
        \address[1]{School of Mathematical and Statistical Sciences, Indian Institute of Technology Mandi\\
        Mandi, Himachal Pradesh, 175005, India}
        
\begin{abstract}
Reconstructing flow fields from sparse and noisy measurements becomes particularly challenging when data are completely absent from dynamically critical regions, such as the vortex-formation region of a wake. Conventional interpolation and numerical reconstruction methods generally require sufficient spatial or boundary information, while physics-informed neural networks (PINNs) can also face difficulties when sparse and noisy observations provide little or no supervision in the region of interest. To address this gap, we propose a Fourier-feature-embedded complex-valued physics-informed convolutional residual network (FCV-PICRN) for sparse, gappy, and noisy flow-field reconstruction. The framework integrates Fourier features for oscillatory and multiscale structures, complex-valued representations for coupled amplitude--phase learning, and convolutional residual blocks constrained by the Navier--Stokes equations. The method is first verified using the Taylor--Green vortex at $Re=100$ and then evaluated for the cylinder wake at $Re=100$. Three severe missing-data configurations are imposed in the dynamically sensitive wake: a large hole, a large truncation, and a plus-shaped gap extending in both streamwise and transverse directions. These gaps are further combined with additive Gaussian and multiplicative speckle noise. Despite having no supervised velocity data inside the excluded regions, FCV-PICRN reconstructs velocity, pressure, and vorticity with high structural similarity index (SSIM) and peak signal-to-noise ratio (PSNR), and remains robust under Gaussian noise up to $20\%$. Finally, for forced two-dimensional Kolmogorov turbulence with $5\%$ Gaussian noise, FCV-PICRN yields lower reconstruction errors than a conventional PINN. These results demonstrate robust physics-informed reconstruction of severely incomplete and noisy flows across laminar and turbulent regimes.
\end{abstract}

\begin{keywords}
Flow-field reconstruction \sep Physics-informed neural networks \sep Complex-valued neural networks \sep Fourier features \sep Gappy and noisy data \sep Turbulent flow
\end{keywords}

	\date{}
	\maketitle
\section{Introduction}
\label{sec:introduction}

Experimental flow measurements are rarely complete and noise free. Particle image velocimetry (PIV) and particle tracking velocimetry (PTV) provide spatially and temporally resolved velocity fields~\cite{adrian1991,raffel2018}, yet model blockage, optical reflections, inadequate seeding, limited field of view, and removal of spurious vectors can leave substantial portions of the measured field unavailable~\cite{westerweel2005}. The consequences are particularly important in bluff-body flows, where missing measurements may overlap the near wake in which separated shear layers roll up and coherent vortices are formed~\cite{williamson1996}. Measurements that remain available are themselves uncertain and may contain signal-dependent noise~\cite{goodman1976}. Thus, a particularly difficult reconstruction problem arises when the region carrying important flow dynamics is precisely the region with little or no reliable measurement information.

Conventional reconstruction methods address different aspects of this problem. Interpolation and robust smoothing are effective for isolated invalid vectors~\cite{garcia2011}, but have limited ability to reconstruct coherent structures that lie entirely inside a large unobserved region. Gappy proper orthogonal decomposition reconstructs missing information from a reduced-order basis~\cite{everson1995,venturi2004}, but depends on representative snapshots from which that basis can be constructed. A more physics-based strategy is to solve the Navier--Stokes equations inside the missing region using measured velocities near its boundary~\cite{sciacchitano2012}. Such an approach, however, requires a dedicated numerical solver and sufficiently reliable boundary information, which becomes increasingly restrictive when the gap is large and the surrounding measurements are noisy. These limitations motivate a reconstruction framework in which the available measurements and governing equations can be exploited simultaneously without requiring a precomputed modal library or an explicit gap-boundary solution.

Physics-informed neural networks (PINNs) provide a natural route toward this objective by incorporating governing equations directly into neural-network training~\cite{lagaris1998,raissi2019,karniadakis2021,cai2021review}. Automatic differentiation supplies the required space--time derivatives~\cite{baydin2018}, allowing physical residuals to be imposed throughout the computational domain, including regions without measurements. This also permits unmeasured quantities, particularly pressure, to be inferred through the governing equations~\cite{raissi2020hfm}. PINNs have consequently been applied to dense flow reconstruction from PIV/PTV measurements~\cite{wang2022dense}, reconstruction from under-resolved and noisy particle tracks~\cite{dileoni2023}, laminar and turbulent incompressible-flow problems~\cite{jin2021nsfnets}, and flow-field super-resolution and denoising~\cite{gao2021}. Bayesian formulations further provide a means of quantifying uncertainty associated with noisy observations~\cite{sun2020bayes,yang2021bpinn}.

Recent studies have moved physics-informed reconstruction toward increasingly incomplete flow measurements. Xu \textit{et al.}~\cite{xu2023practical} demonstrated reconstruction of velocity and pressure in a cylinder wake at $Re=100$ when holes and truncations removed portions of the wake from the training data, although their observations were noise free. Liu \textit{et al.}~\cite{liu2025ffpicn} subsequently developed a feature-fusion physics-informed convolutional network (FFPICN) and reported improved reconstruction for sparse cylinder-wake and isotropic-turbulence data under additive Gaussian noise. Beyond laminar wakes, Parfenyev \textit{et al.}~\cite{parfenyev2024} used PINNs to reconstruct forced two-dimensional turbulence from sparse velocity measurements while identifying unknown physical parameters. These studies establish the capability of physics-informed learning for missing data, noisy measurements, and turbulent flows, but largely address these difficulties separately. A more severe situation arises when a structured gap removes the dynamically sensitive region itself while the remaining observations are simultaneously corrupted by noise. Reconstruction with no supervised velocity measurements inside such gaps, particularly in the presence of both additive Gaussian and multiplicative signal-dependent noise, therefore remains an important challenge.

Addressing this challenge also requires an appropriate representation of the underlying flow. Our recent studies have shown that representation design can substantially influence physics-informed reconstruction. Wavelet-based physics-informed residual neural networks use localized multiscale representations and residual learning for extremely sparse cylinder-wake reconstruction~\cite{barman2026wavelet}, while a lightweight transformer-based PINN uses attention to represent long-range dependencies in incompressible flows~\cite{barman2026physicsformer}. More recently, split complex-valued PINNs demonstrated the potential of complex latent representations for forward and inverse nonlinear partial differential equations~\cite{barman2026scvpinn}. These developments motivate a representation specifically adapted to the oscillatory, multiscale, and phase-dependent structure of unsteady wakes.

Periodic wakes are characterized by a fundamental shedding frequency, its harmonics, and the phase relations among coherent structures. Complex spatial modes and temporal coefficients are therefore widely used to describe such dynamics in modal flow analysis~\cite{schmid2010,taira2017}. Complex-valued neural networks provide a related computational representation because complex arithmetic couples changes in feature amplitude and phase~\cite{hirose2012,trabelsi2018,lee2022survey}. Phase itself contains important structural information in oscillatory signals~\cite{oppenheim1981}. Complex-valued formulations have recently entered physics-informed PDE learning~\cite{si2026complex,zhang2025cvpiml}, including our split complex-valued formulation~\cite{barman2026scvpinn}; however, their use for reconstructing flow fields that are simultaneously sparse, gappy, and noisy remains largely unexplored.

A second representational difficulty is spectral bias. Coordinate-based neural networks tend to learn low-frequency components more readily than higher-frequency structures~\cite{rahaman2019}, which can be restrictive for vortex shedding and broadband turbulent dynamics. Fourier feature mappings enrich coordinate inputs with sinusoidal components and provide an effective mechanism for representing higher-frequency and multiscale behavior~\cite{tancik2020,wang2021eigenvector}. Network architecture provides a complementary consideration: convolutional operations enable compact, weight-shared feature transformations, while residual connections facilitate information propagation and trainability~\cite{he2016resnet,cheng2021deep}. Optimization is also nontrivial because PINN training can suffer from poorly conditioned loss landscapes and competing objectives~\cite{wang2021gradient,krishnapriyan2021}. Recent self-scaled quasi-Newton methods provide an alternative for improving the attainable convergence of physics-informed models~\cite{urban2025}. Together, these observations suggest that frequency representation, amplitude--phase learning, feature processing, physical consistency, and optimization should be treated as connected aspects of the reconstruction problem.

\textbf{Motivation and novelty.}
\textbf{To the best of our knowledge, this is the first study to integrate Fourier-feature embedding, complex-valued representations, and convolutional residual learning within a physics-informed framework specifically for sparse, gappy, and noisy flow-field reconstruction, considering both additive Gaussian and multiplicative speckle noise.} We therefore develop a \textbf{Fourier-Feature-Embedded Complex-Valued Physics-Informed Convolutional Residual Network (FCV-PICRN)}, in which each component addresses a distinct difficulty of the reconstruction. Fourier features enrich the spatiotemporal coordinates to represent oscillatory and multiscale structures and mitigate spectral bias. The embedded coordinates are lifted into a complex-valued latent space so that amplitude and phase can be learned jointly. Complex one-dimensional convolutions provide compact, weight-shared feature transformations, while residual connections facilitate information propagation through the network. The incompressible Navier--Stokes equations constrain the predicted fields throughout the domain and connect the observed flow with regions containing no supervised data. Training combines Adam~\cite{kingma2014adam} with the self-scaled quasi-Newton SSBroyden or SSBFGS method~\cite{urban2025}.

The proposed framework is assessed through progressively more challenging flow-reconstruction settings. First, FCV-PICRN is verified using the Taylor--Green vortex at $Re=100$, where the initial and boundary conditions are imposed and an analytical solution provides a rigorous reference. It is then applied to the cylinder wake at $Re=100$ without separately prescribed initial-, inlet-, outlet-, or wall-condition losses; reconstruction is guided by the available velocity measurements and the incompressible Navier--Stokes residuals. The challenge is increased by introducing large-hole, large-truncation, and plus-shaped gaps in the vortex-formation region together with Gaussian and multiplicative speckle noise. Finally, FCV-PICRN is tested on forced two-dimensional turbulence with sparse velocity data, $5\%$ Gaussian noise, and unknown forcing and dissipation parameters, and is compared with a conventional PINN of comparable size.

The main contributions of this work are summarized as follows:
\begin{enumerate}

    \item \textbf{A unified FCV-PICRN framework for sparse, gappy, and noisy flow reconstruction.}
    Fourier-feature embedding, complex-valued learning, convolutional residual blocks, and the incompressible Navier--Stokes equations are integrated into a single framework for reconstructing flow fields when supervised velocity measurements are unavailable within prescribed regions.

    \item \textbf{A frequency- and phase-aware representation.}
    Fourier features improve the representation of oscillatory and multiscale structures, the complex-valued latent space couples amplitude and phase, and convolutional residual blocks provide compact feature transformation and effective information propagation.

    \item \textbf{Reconstruction under severe structured gaps and distinct noise mechanisms.}
    Large holes, truncations, and plus-shaped gaps are examined together with additive Gaussian and multiplicative speckle noise, including cases in which the dynamically sensitive vortex-formation region contains no supervised velocity measurements.

    \item \textbf{A two-stage physics-informed optimization strategy.}
    Adam is followed by the self-scaled quasi-Newton SSBroyden or SSBFGS method to improve convergence of the proposed physics-informed reconstruction framework.

    \item \textbf{Assessment across laminar and turbulent flow regimes.}
    FCV-PICRN is evaluated from the analytical Taylor--Green vortex to the gappy and noisy cylinder wake and forced two-dimensional turbulence, providing progressively more demanding tests of reconstruction accuracy and robustness.

\end{enumerate}

The remainder of the paper is organized as follows. Section~\ref{sec:preliminary} presents the governing equations and physics-informed formulation. Section~\ref{sec:methodology} describes the FCV-PICRN architecture, observation and gap models, loss formulation, and optimization strategy. Section~\ref{sec:numerical_study} presents the Taylor--Green vortex validation, cylinder-wake reconstruction, and forced two-dimensional turbulence results. Section~\ref{sec:conclusion} summarizes the main conclusions.

\section{Preliminary}
\label{sec:preliminary}

\subsection{Governing Equations}
\label{sec:governing}
We consider an unsteady two-dimensional incompressible flow of a Newtonian fluid over a spatial domain $\Omega\subset\mathbb{R}^{2}$ and a time interval $t\in[0,T]$. Using non-dimensional variables, the governing equations are written as
\begin{align}
  u_t + u\,u_x + v\,u_y + p_x
  - Re^{-1}\left(u_{xx}+u_{yy}\right) &= 0,
  \label{eq:ns_x}\\
  v_t + u\,v_x + v\,v_y + p_y
  - Re^{-1}\left(v_{xx}+v_{yy}\right) &= 0,
  \label{eq:ns_y}\\
  u_x + v_y &= 0,
  \label{eq:continuity}
\end{align}
where $u(x,y,t)$ and $v(x,y,t)$ denote the streamwise and cross-stream velocity components, respectively, and $p(x,y,t)$ represents the kinematic pressure. The Reynolds number is denoted by $Re$, and subscripts indicate partial differentiation with respect to the corresponding variables. The associated scalar vorticity is defined as
\begin{equation}
    \omega = v_x-u_y.
\end{equation}

\subsection{Physics-Informed Neural Networks}
\label{sec:pinn}
\begin{figure}[pos=!htbp]
\refstepcounter{figure}
\begin{center}
    \includegraphics[max width=\textwidth, max height=0.82\textheight]{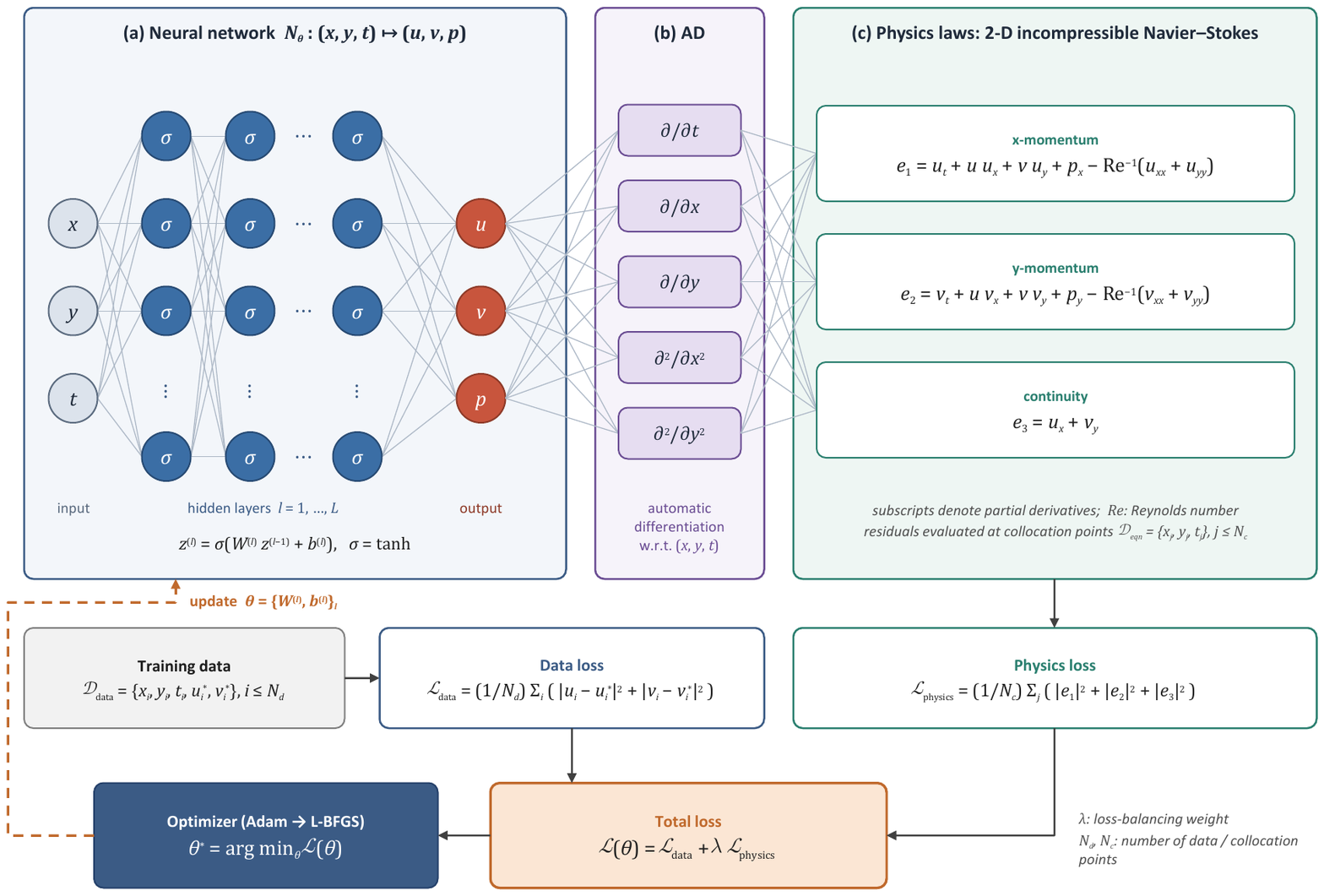}
\end{center}

\begin{quote}
\noindent\textbf{Figure \thefigure.}
Schematic of a conventional PINN for the incompressible Navier--Stokes equations.
\end{quote}
\label{fig1}
\end{figure}

Figure~\ref{fig1} presents the standard PINN architecture~\cite{raissi2019}, which is used as the baseline formulation in this study. The physical variables are approximated by a real-valued fully connected network
$\mathcal{N}_{\theta}:(x,y,t)\mapsto(u,v,p)$, parameterized by $\theta$. For a network containing $L$ hidden layers, the transformation at the $l$th layer is
\begin{equation}
  \mathbf{z}^{(l)}
  =
  \sigma\!\left(
  \mathbf{W}^{(l)}\mathbf{z}^{(l-1)}
  +
  \mathbf{b}^{(l)}
  \right),
  \qquad l=1,\dots,L,
  \label{eq:mlp}
\end{equation}
where $\mathbf{z}^{(0)}=(x,y,t)^{\top}$ denotes the input coordinates, $\mathbf{W}^{(l)}$ and $\mathbf{b}^{(l)}$ are the trainable weight matrix and bias vector of layer $l$, respectively, and $\sigma$ denotes a smooth activation function, commonly $\tanh$. The final affine transformation maps $\mathbf{z}^{(L)}$ to the predicted variables $(u,v,p)$.

The spatial and temporal derivatives required in Equations~\eqref{eq:ns_x}--\eqref{eq:continuity} are evaluated through automatic differentiation. The network parameters are then determined by minimizing a loss that combines the discrepancy from the available velocity measurements with the governing-equation residuals:
\begin{equation}
  \mathcal{L}(\theta)
  =
  \frac{1}{N_d}
  \sum_{i=1}^{N_d}
  \left(
  |u(x_i,y_i,t_i;\theta)-u_i^{*}|^{2}
  +
  |v(x_i,y_i,t_i;\theta)-v_i^{*}|^{2}
  \right)
  +
  \frac{\lambda}{N_c}
  \sum_{j=1}^{N_c}
  \left(
  |e_{1,j}|^{2}
  +
  |e_{2,j}|^{2}
  +
  |e_{3,j}|^{2}
  \right),
  \label{eq:pinn_loss}
\end{equation}
where $(u_i^{*},v_i^{*})$, $i=1,\dots,N_d$, are the measured velocity components at $(x_i,y_i,t_i)$. The quantities $e_1$, $e_2$, and $e_3$ denote the residuals associated with Equations~\eqref{eq:ns_x}--\eqref{eq:continuity}, evaluated at $N_c$ collocation points, and $\lambda$ controls the relative contribution of the physics residual to the data-fitting term. A common optimization strategy employs Adam during the initial training stage followed by L-BFGS for further refinement~\cite{kingma2014adam,liu1989lbfgs}.

For the gappy and noisy reconstruction problem considered in Section~\ref{sec:problem}, this conventional formulation presents several limitations. First, fully connected coordinate networks with smooth non-periodic activations can exhibit spectral bias, making oscillatory and higher-frequency wake structures more difficult to learn~\cite{rahaman2019,wang2021eigenvector}. Second, a purely real-valued representation does not explicitly encode the amplitude--phase interactions associated with periodic flow dynamics. Third, the unscaled data-misfit term in Equation~\eqref{eq:pinn_loss} depends on the magnitude of the measured variables, which can alter its relative balance with the physics residual across problems with different characteristic scales. FCV-PICRN preserves the underlying physics-informed formulation while introducing frequency-aware input features, a complex-valued latent representation, scaled loss terms, and a modified optimization strategy to address these issues.

\subsection{Physical Interpretation: Complex Representation of Fluid Flows}
\label{sec:physical_interpretation}
\begin{figure}[pos=!htbp]
\refstepcounter{figure}
\begin{center}
    \includegraphics[max width=\textwidth, max height=0.82\textheight]{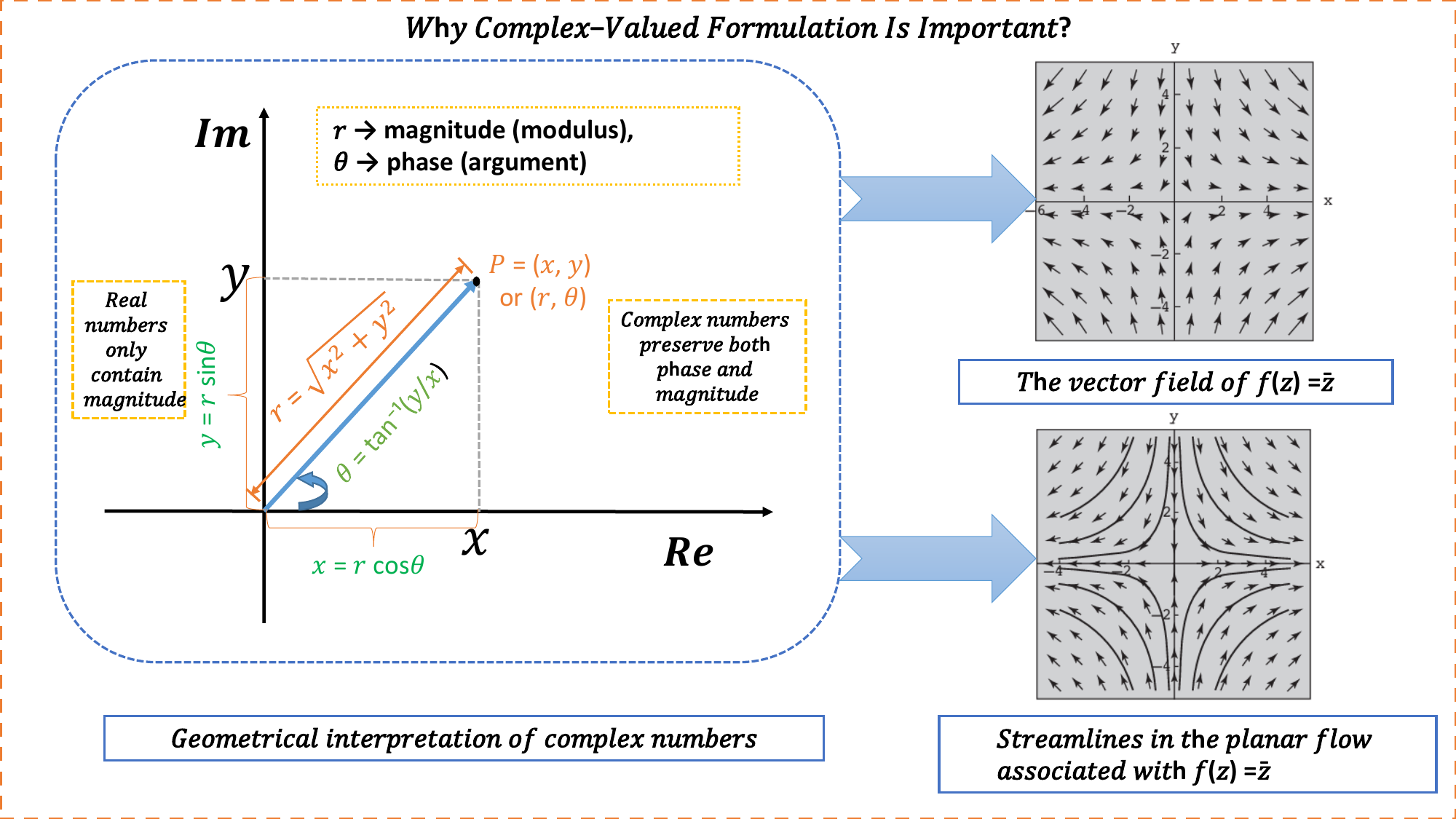}
\end{center}

\begin{quote}
\noindent\textbf{Figure \thefigure.}
Geometric interpretation of a complex number (left) and the planar flow $f(z)=\bar{z}$ shown as its vector field and streamlines (right): a single complex function encodes both the speed and the direction of a two-dimensional flow.
\end{quote}
\label{fig2}
\end{figure}

A complex-valued formulation provides a compact representation of two-dimensional flow quantities by combining magnitude and phase information. For a complex number
$z=x+\mathrm{i}y=r\,e^{\mathrm{i}\chi}$, the modulus
$r=|z|=\sqrt{x^{2}+y^{2}}$ specifies its magnitude and
$\chi=\arg z$ its phase (denoted by $\theta$ in Figure~\ref{fig2}).
This differs from a scalar real-valued representation, as illustrated in
Figure~\ref{fig2}. In the same spirit, a planar velocity field
$\mathbf{u}(x,y)=(u(x,y),v(x,y))$ may be represented as
\begin{equation}
  f(z) = u(x,y) + \mathrm{i}\,v(x,y), \qquad z = x + \mathrm{i}y,
  \label{eq:complex_velocity_field}
\end{equation}
where $\ReP(f)=u$ and $\ImP(f)=v$ correspond to the streamwise and
cross-stream velocity components, respectively. Here,
$\mathrm{i}=\sqrt{-1}$, while $\ReP$ and $\ImP$ denote the real and
imaginary parts of a complex quantity; upright notation is adopted to
distinguish these operators from the Reynolds number $Re$. The modulus
$|f|=\sqrt{u^{2}+v^{2}}$ gives the local speed, whereas $\arg f$ specifies
the local flow direction. Thus, both velocity magnitude and direction can
be expressed through a single complex quantity.

A useful physical analogy follows from two-dimensional incompressible and
irrotational flow, for which the complex potential is written as
\begin{equation}
  F(z) = \phi(x,y) + \mathrm{i}\,\psi(x,y),
  \label{eq:complex_potential}
\end{equation}
with $\phi(x,y)$ and $\psi(x,y)$ denoting the velocity potential and stream
function, respectively. Analyticity of $F$ gives the Cauchy--Riemann
relations
\begin{equation}
  u = \frac{\partial\phi}{\partial x}
    = \frac{\partial\psi}{\partial y}, \qquad
  v = \frac{\partial\phi}{\partial y}
    = -\frac{\partial\psi}{\partial x},
  \label{eq:cauchy_riemann}
\end{equation}
which satisfy $u_x+v_y=0$ and $v_x-u_y=0$. The corresponding complex
velocity is $\mathrm{d}F/\mathrm{d}z=u-\mathrm{i}v$, and hence
$f(z)=\overline{\mathrm{d}F/\mathrm{d}z}$. The level curves
$\psi(x,y)=\text{constant}$ define the streamlines and, for steady flow,
coincide with fluid-particle trajectories.

Figure~\ref{fig2} demonstrates this relation using
$F(z)=z^{2}/2$. Since $\mathrm{d}F/\mathrm{d}z=z$, one obtains
$f(z)=\bar{z}=x-\mathrm{i}y$, or equivalently $u=x$ and $v=-y$.
This corresponds to a planar stagnation-point flow with streamlines
$\psi=xy=\text{constant}$. The example illustrates how two velocity
components can be expressed through the real and imaginary parts of a
single complex representation, with their relation governed by
Equation~\eqref{eq:cauchy_riemann}. It therefore provides a simple
interpretation of how magnitude, direction, and phase-related information
can be organized within a complex-valued description.

The viscous, rotational, and unsteady cylinder wake considered in this
work is not a potential flow; consequently, the analytic relations in
Equation~\eqref{eq:cauchy_riemann} are not imposed on the reconstructed
fields. Instead, the velocity components are coupled through the
continuity and momentum equations,
Equations~\eqref{eq:ns_x}--\eqref{eq:continuity}, while periodic vortex
shedding introduces strong temporal and spatial phase relationships.
This motivates the use of a complex-valued latent representation rather
than treating the potential-flow analogy as a physical constraint.
In FCV-PICRN, amplitude and phase information can interact within the
complex hidden features (Section~\ref{sec:complex}), whereas the physical
variables are obtained from the real parts of the complex outputs
(Equation~\eqref{eq:real_parts}). The complex formulation therefore serves
as a representation mechanism for learning coupled flow structures, rather
than assuming analyticity of the underlying viscous flow. This viewpoint
is consistent with the motivation of our earlier split complex-valued
PINN~\cite{barman2026scvpinn} and provides the basis for extending
complex-valued physics-informed learning to the present gappy and noisy
flow-reconstruction problem.

\section{Methodology}
\label{sec:methodology}
\textbf{Motivation and overview.}
Reconstructing a flow field from gappy and noisy measurements requires a model that can represent oscillatory structures, remain stable during optimization, and enforce physical consistency where observations are unavailable. FCV-PICRN addresses these requirements through complementary components: Fourier features enrich the coordinate representation with oscillatory content; a complex-valued latent space enables coupled amplitude--phase learning; complex convolutions and residual connections provide compact feature transformation and effective information propagation; and the Navier--Stokes residuals constrain the reconstruction within both observed and unobserved regions. The following sections describe the reconstruction problem (Section~\ref{sec:problem}), gap and noise models (Sections~\ref{sec:gaps}--\ref{sec:noise}), network architecture (Section~\ref{sec:architecture}), physics-informed formulation (Sections~\ref{sec:residuals}--\ref{sec:loss}), optimization strategy (Section~\ref{sec:optimization}), and evaluation metrics (Section~\ref{sec:metrics}). Unless stated otherwise, the same network, loss formulation, and optimization procedure are used throughout the test cases.

\subsection{Reconstruction Problem}
\label{sec:problem}
We consider the two-dimensional incompressible flow described by
Equations~\eqref{eq:ns_x}--\eqref{eq:continuity}. The available measurements are represented by
\begin{equation}
  \mathcal{D}_{\mathrm{data}}
  =
  \left\{
  \left(x_i,y_i,t_i;\;u_i^{*},v_i^{*}\right)
  \right\}_{i=1}^{N_d},
  \label{eq:data_set}
\end{equation}
where $u_i^{*}$ and $v_i^{*}$ denote the measured, and possibly noise-corrupted, velocity components at $(x_i,y_i,t_i)$. No velocity observations are provided within a prescribed gap $\Omega_{\mathrm{hole}}\subset\Omega$, and pressure measurements are not used. The objective is to reconstruct $u$, $v$, $p$, and $\omega$ over $\Omega\times[0,T]$, including the unobserved region $\Omega_{\mathrm{hole}}$, while remaining consistent with the available velocity data and Equations~\eqref{eq:ns_x}--\eqref{eq:continuity}. Since $p$ enters Equations~\eqref{eq:ns_x}--\eqref{eq:ns_y} only through its spatial gradient and is not directly supervised, it is identifiable only up to an additive function of time. The reconstructed pressure is therefore compared with the reference field after removing this offset~\cite{xu2023practical}.

For the present reconstruction setting, no separate boundary or initial conditions are imposed. In particular, the cylinder-wake domain is extracted as a window from a larger flow field, so the velocities along its boundaries are not prescribed independently of the measurements. Instead, reconstruction is guided by the observations distributed over
$\Omega\setminus\Omega_{\mathrm{hole}}$ and $[0,T]$, together with the governing-equation constraints. When boundary or initial information is available for a given problem, the corresponding penalty terms can be incorporated into the loss in Section~\ref{sec:loss} without altering the overall FCV-PICRN formulation.

\subsection{Gappy Observation Model}
\label{sec:gaps}
To describe different missing-data geometries within a common formulation, the gap is constructed from $K_h$ sub-regions. Each sub-region is specified by a signed shape function $g_k(x,y)$, with $g_k(x,y)\leq0$ identifying points inside the corresponding exclusion:
\begin{equation}
  \Omega_{\mathrm{hole}} = \bigcup_{k=1}^{K_h}\left\{(x,y)\in\Omega : g_k(x,y)\le 0\right\}.
  \label{eq:hole_union}
\end{equation}
The geometries considered here are defined by
\begin{align}
  g_k^{\mathrm{rect}}(x,y) &= \max\!\left(\frac{|x-x_c^{(k)}|}{a^{(k)}},\;\frac{|y-y_c^{(k)}|}{b^{(k)}}\right) - 1, \label{eq:g_rect}\\
  g_k^{\mathrm{ell}}(x,y) &= \left(\frac{x-x_c^{(k)}}{a^{(k)}}\right)^{2} + \left(\frac{y-y_c^{(k)}}{b^{(k)}}\right)^{2} - 1, \label{eq:g_ell}\\
  g_k^{\mathrm{cross}}(x,y) &= \min\!\left(g_{k,\mathrm{vert}}^{\mathrm{rect}}(x,y),\; g_{k,\mathrm{horz}}^{\mathrm{rect}}(x,y)\right), \label{eq:g_cross}
\end{align}
where $(x_c^{(k)},y_c^{(k)})$ denotes the center of the $k$th sub-region, and $a^{(k)}$ and $b^{(k)}$ are its half-widths along the $x$ and $y$ directions, respectively. Equation~\eqref{eq:g_rect} defines a rectangular gap; taking $a^{(k)}\to\infty$ or $b^{(k)}\to\infty$ gives a strip extending across the full width or height of the domain. Equation~\eqref{eq:g_ell} represents an elliptical gap and reduces to a circular one when $a^{(k)}=b^{(k)}$. The cross-shaped gap in Equation~\eqref{eq:g_cross} is formed by the union of vertical and horizontal rectangular strips, with the minimum operator selecting points that belong to either strip. The general definition in Equation~\eqref{eq:hole_union} also permits polygonal or irregular masks without altering the remaining formulation.

For each numerical case in Section~\ref{sec:numerical_study}, supervised samples are selected from the prescribed reference snapshots or measurement points. Any sample located within $\Omega_{\mathrm{hole}}$ is removed before training, ensuring that no supervised data are available inside the designated gap.

\subsection{Measurement-Noise Models}
\label{sec:noise}
The retained velocity observations are perturbed using two noise models with noise level $\eta$. Gaussian noise is considered at $\eta=5\%$, $10\%$, $15\%$, and $20\%$, whereas speckle noise is examined at $\eta=5\%$ (Section~\ref{sec:numerical_study}). Noise realizations are generated independently for each observation and for the two velocity components. Here, $u_i$ and $v_i$ denote the clean reference velocities at $(x_i,y_i,t_i)$.

\textbf{Model A: additive Gaussian noise.}
The Gaussian model introduces an additive perturbation scaled by the global standard deviation of each velocity component:
\begin{equation}
  u_i^{*} = u_i + \eta\,\mathrm{std}(u)\,\varepsilon_i, \qquad
  v_i^{*} = v_i + \eta\,\mathrm{std}(v)\,\varepsilon_i', \qquad
  \varepsilon_i,\,\varepsilon_i' \sim \mathcal{N}(0,1),
  \label{eq:noise_gauss}
\end{equation}
where $\mathrm{std}(\cdot)$ denotes the standard deviation of the clean retained samples of the corresponding component. Since the perturbation amplitude is independent of the local velocity value, Equation~\eqref{eq:noise_gauss} represents a spatially uniform noise level relative to the global variation of each component.

\textbf{Model B: multiplicative speckle noise.}
To examine signal-dependent corruption, we adopt the multiplicative speckle model used in image processing and implemented by the \texttt{random\_noise} function with mode \texttt{`speckle'} in scikit-image~\cite{vanderwalt2014}:
\begin{equation}
  u_i^{*} = u_i + u_i\,\varepsilon_i = u_i\left(1+\varepsilon_i\right), \qquad
  v_i^{*} = v_i\left(1+\varepsilon_i'\right), \qquad
  \varepsilon_i,\,\varepsilon_i' \sim \mathcal{N}\!\left(0,\eta^{2}\right).
  \label{eq:noise_speckle}
\end{equation}
Unlike Model A, the perturbation magnitude now depends on the local velocity. In particular,
$\mathbb{E}[u_i^{*}]=u_i$ and the corresponding standard deviation is
$\eta\,|u_i|$. The resulting corruption is therefore spatially heterogeneous: it increases with the magnitude of the velocity component and approaches zero where that component vanishes. This behavior is representative of signal-dependent errors encountered in coherent and image-based measurements~\cite{goodman1976} and provides a complementary robustness test to the additive Gaussian model.

For both noise models, reconstruction errors are evaluated against the clean reference fields (Section~\ref{sec:metrics}), thereby measuring the ability of the proposed framework to reconstruct the underlying flow rather than reproduce the imposed measurement perturbations.

\subsection{Proposed FCV-PICRN Architecture}
\label{sec:architecture}

\begin{figure}[pos=!htbp]
\refstepcounter{figure}
\begin{center}
    \includegraphics[max width=\textwidth, max height=0.82\textheight]{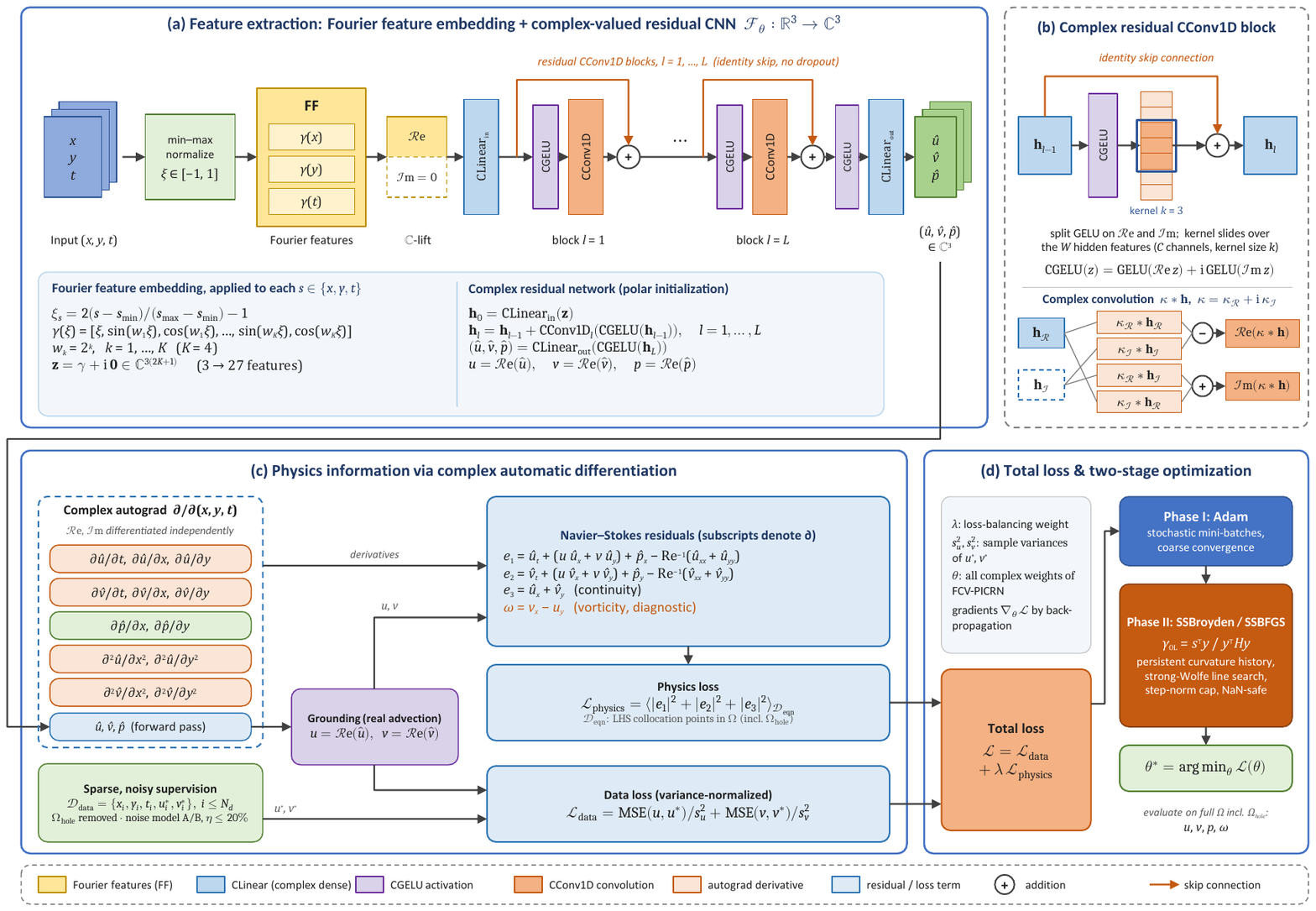}
\end{center}

\begin{quote}
\noindent\textbf{Figure \thefigure.}
Architecture of the proposed FCV-PICRN, comprising (a) the Fourier-embedded complex convolutional-residual network $\mathcal{F}_{\theta}$, (b) the complex residual block, (c) the physics-informed losses, and (d) the two-stage optimization.
\end{quote}
\label{fig3}
\end{figure}

Figure~\ref{fig3} shows the proposed FCV-PICRN architecture, defined by the mapping
\begin{equation}
  \mathcal{F}_{\theta}:\;(x,y,t)\in\mathbb{R}^{3}\;\longmapsto\;(\hat{u},\hat{v},\hat{p})\in\mathbb{C}^{3},
  \label{eq:network_map}
\end{equation}
where $\theta$ denotes the trainable parameters and a hat indicates a network prediction. The architecture consists of a real Fourier feature embedding, a complex input layer, $L$ complex convolutional residual blocks, and a complex output layer. The physical variables are obtained from the real parts of the complex outputs,
\begin{equation}
  u=\ReP(\hat{u}),\qquad v=\ReP(\hat{v}),\qquad p=\ReP(\hat{p}).
  \label{eq:real_parts}
\end{equation}
As in a conventional PINN, $\mathcal{F}_{\theta}$ is evaluated pointwise at each space--time coordinate. The individual components are described below.

\subsubsection{Fourier Feature Embedding}
\label{sec:fourier}

Coordinate-based networks tend to learn low-frequency components before higher-frequency ones, a behavior commonly associated with spectral bias~\cite{rahaman2019}. In the neural tangent kernel interpretation, the error component associated with the $j$th kernel eigenfunction evolves approximately as
\begin{equation}
  \delta_j(n) \approx \left(1-\tau\,\mu_j\right)^{n}\delta_j(0),
  \label{eq:ntk_decay}
\end{equation}
where $n$ denotes the iteration, $\tau$ the learning rate, and $\mu_j$ the corresponding eigenvalue~\cite{wang2021eigenvector}. Since the eigenvalues of coordinate networks generally decrease with increasing frequency, oscillatory components can converge more slowly. This is relevant to wake flows containing a shedding frequency, higher harmonics, and convected vortical structures. Fourier features provide these oscillatory components directly at the input level~\cite{tancik2020,wang2021eigenvector}.

Each coordinate $s\in\{x,y,t\}$ is first normalized to $[-1,1]$,
\begin{equation}
  \xi_s = 2\,\frac{s-s_{\min}}{s_{\max}-s_{\min}} - 1,
  \label{eq:minmax}
\end{equation}
and subsequently embedded as
\begin{equation}
  \gamma(\xi) = \left[\,\xi,\;\sin(w_1\xi),\;\cos(w_1\xi),\;\dots,\;\sin(w_K\xi),\;\cos(w_K\xi)\,\right],
  \qquad w_k = 2^{k},\quad k=1,\dots,K.
  \label{eq:fourier}
\end{equation}
Applying Equation~\eqref{eq:fourier} to $\xi_x$, $\xi_y$, and $\xi_t$ and concatenating the resulting features gives
$\boldsymbol{\gamma}\in\mathbb{R}^{3(2K+1)}$. For the default $K=4$, the embedding contains 27 features. For the cylinder wake, the frequencies are shifted by one octave to $w_k=2^{k-1}$, giving $1,2,4,8$. These frequencies are fixed, axis aligned, and logarithmically spaced, similar to positional encoding in coordinate networks~\cite{mildenhall2020}, rather than randomly sampled as in Gaussian Fourier feature mappings~\cite{tancik2020}.

For a single coordinate, the first trainable layer forms combinations of the embedded features of the form
\begin{equation}
  a_0\,\xi + \sum_{k=1}^{K}\left[a_k\sin(w_k\xi) + b_k\cos(w_k\xi)\right],
  \label{eq:fourier_series}
\end{equation}
where $a_k$ and $b_k$ are trainable coefficients. Thus, the network begins with an explicit oscillatory basis instead of constructing it entirely through subsequent nonlinear transformations. Interactions among spatial and temporal features can further generate travelling-wave components. For example,
\begin{equation}
  \sin(w_j\xi_x)\cos(w_k\xi_t) = \tfrac{1}{2}\left[\sin(w_j\xi_x + w_k\xi_t) + \sin(w_j\xi_x - w_k\xi_t)\right],
  \label{eq:product_to_sum}
\end{equation}
which contains combinations analogous to $\sin(\kappa x-\varpi t)$ and is therefore suitable for representing convected periodic structures.

The embedding also contributes directly to the derivatives entering the physics residuals:
\begin{equation}
  \frac{\partial}{\partial s}\sin(w_k\xi_s) = \frac{2w_k}{s_{\max}-s_{\min}}\cos(w_k\xi_s), \qquad
  \frac{\partial^2}{\partial s^2}\sin(w_k\xi_s) = -\left(\frac{2w_k}{s_{\max}-s_{\min}}\right)^{2}\sin(w_k\xi_s).
  \label{eq:fourier_derivatives}
\end{equation}
The first- and second-order derivatives therefore retain frequency-dependent contributions required by the advective and viscous terms in Equations~\eqref{eq:ns_x}--\eqref{eq:ns_y}. The selected frequencies span normalized wavelengths from $2\pi/w_K$ to $2\pi/w_1$ and can be adjusted for flows with different spectral characteristics.

A fixed Fourier embedding is used instead of applying periodic activations throughout the complex network. For a complex argument,
$\sin(a+\mathrm{i}b)=\sin a\cosh b+\mathrm{i}\cos a\sinh b$, so the hyperbolic terms can grow rapidly with $|b|$ and become numerically problematic during large quasi-Newton updates. In contrast, Equation~\eqref{eq:fourier} is evaluated only once on bounded real coordinates. The resulting features are lifted to the complex space with zero imaginary part:
\begin{equation}
  \mathbf{z} = \boldsymbol{\gamma} + \mathrm{i}\,\mathbf{0}\in\mathbb{C}^{3(2K+1)}.
  \label{eq:complex_lift}
\end{equation}

\subsubsection{Complex-Valued Representation}
\label{sec:complex}

Periodic flow structures can be expressed naturally through amplitude and phase. For example, a single shedding mode may be written as
\begin{equation}
  q(\mathbf{x},t)=\ReP\!\left\{q_{\mathrm{c}}(\mathbf{x})\,e^{\mathrm{i}\varpi t}\right\},\qquad q_{\mathrm{c}}(\mathbf{x})=A(\mathbf{x})\,e^{\mathrm{i}\varphi(\mathbf{x})},
  \label{eq:mode}
\end{equation}
where $\varpi$ is the angular frequency, $A$ is the local amplitude, and $\varphi$ is the phase. Complex modes are widely used to characterize periodic and coherent flow dynamics~\cite{schmid2010,taira2017}. A phase shift can then be represented directly through multiplication by a unit complex number, motivating the use of complex-valued latent features~\cite{hirose2012,trabelsi2018,lee2022survey}.

All trainable layers following the Fourier embedding operate in the complex domain. Let
$\mathbf{W}=\mathbf{W}_{\mathcal{R}}+\mathrm{i}\mathbf{W}_{\mathcal{I}}$
and
$\mathbf{z}=\mathbf{z}_{\mathcal{R}}+\mathrm{i}\mathbf{z}_{\mathcal{I}}$,
where
$\mathbf{W}_{\mathcal{R}}=\ReP(\mathbf{W})$,
$\mathbf{W}_{\mathcal{I}}=\ImP(\mathbf{W})$, and similarly for $\mathbf{z}$. Their product is
\begin{equation}
  \mathbf{W}\mathbf{z} = \left(\mathbf{W}_{\mathcal{R}}\mathbf{z}_{\mathcal{R}}-\mathbf{W}_{\mathcal{I}}\mathbf{z}_{\mathcal{I}}\right)
  + \mathrm{i}\left(\mathbf{W}_{\mathcal{R}}\mathbf{z}_{\mathcal{I}}+\mathbf{W}_{\mathcal{I}}\mathbf{z}_{\mathcal{R}}\right),
  \label{eq:complex_product}
\end{equation}
which can be evaluated using real-valued operations~\cite{trabelsi2018}. For a single weight $w=a+\mathrm{i}b$ acting on $z=z_{\mathcal{R}}+\mathrm{i}z_{\mathcal{I}}$, the same operation becomes
\begin{equation}
  \begin{pmatrix} z_{\mathcal{R}}' \\ z_{\mathcal{I}}' \end{pmatrix}
  = \begin{pmatrix} a & -b \\ b & a \end{pmatrix}
    \begin{pmatrix} z_{\mathcal{R}} \\ z_{\mathcal{I}} \end{pmatrix}
  = |w|\begin{pmatrix} \cos\vartheta & -\sin\vartheta \\ \sin\vartheta & \cos\vartheta \end{pmatrix}
    \begin{pmatrix} z_{\mathcal{R}} \\ z_{\mathcal{I}} \end{pmatrix},
  \qquad \vartheta=\arg w.
  \label{eq:rotation_scaling}
\end{equation}
Hence, a complex weight combines scaling by $|w|$ with rotation by $\vartheta$. For $z=Ae^{\mathrm{i}\varphi}$,
$wz=|w|A\,e^{\mathrm{i}(\varphi+\vartheta)}$, so amplitude and phase are transformed jointly. A complex layer of width $W$ contains $2W^{2}$ real weight parameters, compared with $4W^{2}$ parameters for an unconstrained real $2W\times2W$ transformation. The complex formulation therefore imposes the rotation--scaling structure of Equation~\eqref{eq:rotation_scaling} while retaining a compact parameterization. Since phase contains substantial structural information in oscillatory signals~\cite{oppenheim1981}, this representation is well suited to coherent wake dynamics.

The network is not assumed to be holomorphic. All derivatives required by the governing equations are taken with respect to the real coordinates $(x,y,t)$, with the real and imaginary output components differentiated separately using automatic differentiation. Consequently, no Cauchy--Riemann constraint is imposed on the network activation.

Complex weights are initialized in polar form following Trabelsi \textit{et al.}~\cite{trabelsi2018}. The magnitude $|w|$ is sampled from a Rayleigh distribution with scale $s$, while $\arg w$ is sampled uniformly from $[-\pi,\pi]$. For the Taylor--Green vortex and turbulent-flow cases,
$s=\sqrt{2/(n_{\mathrm{in}}+n_{\mathrm{out}})}$, whereas for the cylinder wake
$s=1/\sqrt{n_{\mathrm{in}}+n_{\mathrm{out}}}$, where $n_{\mathrm{in}}$ and $n_{\mathrm{out}}$ are the layer fan-in and fan-out, respectively. The latter gives
$\mathbb{E}|w|^{2}=2s^{2}=2/(n_{\mathrm{in}}+n_{\mathrm{out}})$, corresponding to the complex Glorot criterion.

\subsubsection{Complex Convolutional Layers}
\label{sec:layers}

Sparse and noisy supervision favors a compact parameterization that limits unnecessary degrees of freedom while retaining sufficient representation capacity. Convolutional layers achieve this through weight sharing and have also been used successfully in physics-informed flow reconstruction~\cite{gao2021,liu2025ffpicn}.

Dense complex layers are retained only for input lifting and output projection:
\begin{equation}
  \mathrm{CLinear}(\mathbf{z}) = \mathbf{W}\mathbf{z}+\mathbf{b},\qquad \mathbf{W}\in\mathbb{C}^{n_{\mathrm{out}}\times n_{\mathrm{in}}},\;\mathbf{b}\in\mathbb{C}^{n_{\mathrm{out}}},
  \label{eq:clinear}
\end{equation}
with
$\mathrm{CLinear}_{\mathrm{in}}:\mathbb{C}^{3(2K+1)}\to\mathbb{C}^{W}$
and
$\mathrm{CLinear}_{\mathrm{out}}:\mathbb{C}^{W}\to\mathbb{C}^{3}$,
where $W$ denotes the hidden width. The intermediate transformations use complex one-dimensional convolutions. For the hidden feature vector
$\mathbf{h}=(h_1,\dots,h_W)^{\top}\in\mathbb{C}^{W}$, block $l$ computes
\begin{equation}
  \left[\mathrm{CConv1D}_l(\mathbf{h})\right]_j = \sum_{m=-r}^{r}\kappa_{l,m}\,h_{j+m} + \beta_l,
  \qquad j=1,\dots,W,
  \label{eq:cconv}
\end{equation}
where $\kappa_{l,m}\in\mathbb{C}$ are shared kernel weights, $\beta_l\in\mathbb{C}$ is the bias, and the kernel size is $k=2r+1$ ($k=3$ by default). Zero padding, $h_{j+m}=0$ outside $\{1,\dots,W\}$, preserves the feature length.

Equation~\eqref{eq:cconv} can equivalently be expressed as
$\mathrm{CConv1D}_l(\mathbf{h})=\mathbf{T}_l\mathbf{h}+\beta_l\mathbf{1}$,
where, for $k=3$,
\begin{equation}
  \mathbf{T}_l=
  \begin{pmatrix}
    \kappa_{l,0} & \kappa_{l,1} & & & \\
    \kappa_{l,-1} & \kappa_{l,0} & \kappa_{l,1} & & \\
     & \ddots & \ddots & \ddots & \\
     & & \kappa_{l,-1} & \kappa_{l,0} & \kappa_{l,1}\\
     & & & \kappa_{l,-1} & \kappa_{l,0}
  \end{pmatrix}\in\mathbb{C}^{W\times W}.
  \label{eq:toeplitz}
\end{equation}
Unlike a dense $\mathbb{C}^{W\times W}$ transformation with $W^{2}$ independent complex weights, the convolution uses only $k$ shared kernel coefficients. For example, with $K=4$, $L=4$, $k=3$, and $W=64$, the complete network contains 4,006 real parameters: 3,584 in $\mathrm{CLinear}_{\mathrm{in}}$, 32 in the four convolutional blocks, and 390 in $\mathrm{CLinear}_{\mathrm{out}}$. By comparison, a single dense $64\times64$ complex hidden layer requires 8,320 real parameters.

More generally, $\mathbf{h}$ may be organized into $C$ channels of length $W/C$. The kernel coefficients then become
$\boldsymbol{\kappa}_{l,m}\in\mathbb{C}^{C\times C}$,
the bias becomes $\boldsymbol{\beta}_l\in\mathbb{C}^{C}$, and each block contains
$2(kC^{2}+C)$ real parameters. Equation~\eqref{eq:cconv} corresponds to the single-channel case $C=1$. The width and channel configuration are selected for each flow and reported in Section~\ref{sec:numerical_study}.

A single convolution couples neighboring features, while stacking $L$ blocks increases the effective coupling range to $L(k-1)+1$ features. The dense input and output layers provide global mixing at the two ends of the network. Importantly, these convolutions operate along the latent feature dimension rather than over a physical spatial grid. FCV-PICRN therefore remains a pointwise differentiable mapping of $(x,y,t)$, allowing the governing-equation residuals in Section~\ref{sec:residuals} to be evaluated at arbitrary space--time coordinates.

\subsubsection{Residual Connections}
\label{sec:assembly}

Residual connections are introduced to improve information and gradient propagation through the stacked convolutional blocks, which is particularly important when first- and second-order coordinate derivatives are required by the physics-informed loss.

Using the complex embedded vector $\mathbf{z}$ from Equation~\eqref{eq:complex_lift}, the network is assembled as
\begin{align}
  \mathbf{h}_0 &= \mathrm{CLinear}_{\mathrm{in}}(\mathbf{z}), \label{eq:res_in}\\
  \mathbf{h}_l &= \mathbf{h}_{l-1} + \mathcal{B}_l(\mathbf{h}_{l-1}),\qquad \mathcal{B}_l(\mathbf{h})=\mathrm{CConv1D}_l\!\left(\mathrm{CGELU}(\mathbf{h})\right), \qquad l=1,\dots,L, \label{eq:res_block}\\
  (\hat{u},\hat{v},\hat{p})^{\top} &= \mathrm{CLinear}_{\mathrm{out}}\!\left(\mathrm{CGELU}(\mathbf{h}_L)\right), \label{eq:res_out}
\end{align}
where CGELU is defined in Section~\ref{sec:activation}, and the physical outputs follow Equation~\eqref{eq:real_parts}. Each residual block adds the learned correction $\mathcal{B}_l$ to an identity path~\cite{he2016resnet}. The Jacobian across the stack is
\begin{equation}
  \frac{\partial\mathbf{h}_L}{\partial\mathbf{h}_0}=\prod_{l=L}^{1}\left(\mathbf{I}+\mathbf{J}_l\right)
  = \mathbf{I}+\sum_{l=1}^{L}\mathbf{J}_l+\cdots,
  \qquad \mathbf{J}_l=\frac{\partial\mathcal{B}_l}{\partial\mathbf{h}_{l-1}}.
  \label{eq:res_jacobian}
\end{equation}
The identity contribution provides a direct propagation path through the network and reduces the degradation of gradients associated with repeatedly composed nonlinear transformations. The same Jacobian enters coordinate differentiation,
$\partial\mathbf{h}_L/\partial s=
(\partial\mathbf{h}_L/\partial\mathbf{h}_0)
(\partial\mathbf{h}_0/\partial s)$,
which facilitates propagation of the Fourier-feature derivatives in Equation~\eqref{eq:fourier_derivatives} to the output fields and hence to the advective and viscous residual terms.

When the residual corrections $\mathcal{B}_l$ are small, the architecture remains close to a shallow Fourier-feature model, while the stacked blocks add representation capacity as required by the data and governing equations. Residual architectures have previously been employed in PINNs for fluid-flow problems~\cite{cheng2021deep,barman2026wavelet}. No dropout or normalization layers are used, preserving a deterministic and differentiable mapping from the coordinates to the predicted fields.

\subsubsection{Activation Function}
\label{sec:activation}

The activation function must remain sufficiently smooth for the second-order derivatives appearing in the viscous terms while maintaining numerically stable growth during the quasi-Newton optimization stage. We therefore use the split complex Gaussian error linear unit (CGELU)~\cite{hendrycks2016},
\begin{equation}
  \mathrm{CGELU}(z) = \mathrm{GELU}\!\left(\ReP(z)\right) + \mathrm{i}\,\mathrm{GELU}\!\left(\ImP(z)\right),
  \qquad \mathrm{GELU}(a) = a\,\Phi(a),
  \label{eq:cgelu}
\end{equation}
where $\Phi$ denotes the standard normal cumulative distribution function. The split formulation applies the real-valued activation independently to the real and imaginary components, consistent with common complex-valued network formulations~\cite{trabelsi2018} and our earlier split complex-valued PINN~\cite{barman2026scvpinn}.

GELU is smooth, possesses nonzero second derivatives over its nonlinear transition region, and grows at most linearly. These properties make it suitable for physics-informed problems containing second-order differential operators. In contrast, a split ReLU produces a piecewise-linear mapping whose second derivative vanishes almost everywhere, making it unsuitable for accurately representing the viscous terms through automatic differentiation; related limitations of ReLU in Navier--Stokes PINNs have also been discussed by Xu \textit{et al.}~\cite{xu2023practical}. Complex sine activations are not adopted here because of the potentially rapid growth of their hyperbolic components for complex arguments, as discussed in Section~\ref{sec:fourier}.

\subsection{Physics-Informed Residuals}
\label{sec:residuals}
\textit{Why they are needed.}
Within $\Omega_{\mathrm{hole}}$, no supervised measurements are available; hence, the governing equations provide the primary constraint on the reconstructed fields. They also regularize the solution against measurement noise, since arbitrary perturbations are generally inconsistent with the flow equations, and enable pressure to be inferred without direct pressure measurements.

\textit{How they are formed.}
The residuals corresponding to Equations~\eqref{eq:ns_x}--\eqref{eq:continuity} are constructed from the complex network outputs. The temporal, pressure-gradient, and viscous terms act on the complex fields, whereas the nonlinear transport terms use the real physical velocities defined in Equation~\eqref{eq:real_parts}:
\begin{align}
  e_1 &= \hat{u}_t + u\,\hat{u}_x + v\,\hat{u}_y + \hat{p}_x - Re^{-1}\left(\hat{u}_{xx}+\hat{u}_{yy}\right), \label{eq:e1}\\
  e_2 &= \hat{v}_t + u\,\hat{v}_x + v\,\hat{v}_y + \hat{p}_y - Re^{-1}\left(\hat{v}_{xx}+\hat{v}_{yy}\right), \label{eq:e2}\\
  e_3 &= \hat{u}_x + \hat{v}_y. \label{eq:e3}
\end{align}
All required derivatives are evaluated by automatic differentiation through the complex network, with the real and imaginary components differentiated separately as described in Section~\ref{sec:complex}. The vorticity,
$\omega=v_x-u_y$, is obtained from the same derivatives and is used only for evaluating the reconstructed flow.

To clarify the role of the complex formulation, write
$\hat{u}=u+\mathrm{i}\,u^{\mathcal{I}}$,
$\hat{v}=v+\mathrm{i}\,v^{\mathcal{I}}$, and
$\hat{p}=p+\mathrm{i}\,p^{\mathcal{I}}$, where
$u^{\mathcal{I}}\equiv\ImP(\hat{u})$,
$v^{\mathcal{I}}\equiv\ImP(\hat{v})$, and
$p^{\mathcal{I}}\equiv\ImP(\hat{p})$.
Since differentiation with respect to the real coordinates commutes with the real--imaginary decomposition, e.g.,
$\ReP(\hat{u}_x)=u_x$ and
$\ImP(\hat{u}_x)=u_x^{\mathcal{I}}$, the residuals separate exactly into real and imaginary components. Their real parts satisfy
\begin{equation}
  \ReP(e_k) = r_k(u,v,p),\qquad k=1,2,3,
  \label{eq:split_re}
\end{equation}
where $r_1$, $r_2$, and $r_3$ are the left-hand sides of Equations~\eqref{eq:ns_x}, \eqref{eq:ns_y}, and \eqref{eq:continuity}, respectively, evaluated using the physical fields $(u,v,p)$. The corresponding imaginary parts are
\begin{align}
  \ImP(e_1) &= u^{\mathcal{I}}_t + u\,u^{\mathcal{I}}_x + v\,u^{\mathcal{I}}_y + p^{\mathcal{I}}_x - Re^{-1}\left(u^{\mathcal{I}}_{xx}+u^{\mathcal{I}}_{yy}\right), \label{eq:split_e1}\\
  \ImP(e_2) &= v^{\mathcal{I}}_t + u\,v^{\mathcal{I}}_x + v\,v^{\mathcal{I}}_y + p^{\mathcal{I}}_y - Re^{-1}\left(v^{\mathcal{I}}_{xx}+v^{\mathcal{I}}_{yy}\right), \label{eq:split_e2}\\
  \ImP(e_3) &= u^{\mathcal{I}}_x + v^{\mathcal{I}}_y. \label{eq:split_e3}
\end{align}

Because
\begin{equation}
  |e_k|^{2}
  =
  \left[\ReP(e_k)\right]^{2}
  +
  \left[\ImP(e_k)\right]^{2},
\end{equation}
the complex residual penalty constrains both components simultaneously. Equation~\eqref{eq:split_re} shows that the real component is exactly the governing-equation residual of the reconstructed physical fields; therefore, minimizing it directly enforces Equations~\eqref{eq:ns_x}--\eqref{eq:continuity}. The imaginary components in Equations~\eqref{eq:split_e1}--\eqref{eq:split_e3} form an auxiliary linear transport--diffusion system driven by the reconstructed velocity $(u,v)$ and governed by the same Reynolds number.

The imaginary variables do not enter the data loss and do not modify the physical equations directly. In particular, $(u^{\mathcal{I}},v^{\mathcal{I}},p^{\mathcal{I}})=(0,0,0)$ satisfies the auxiliary system. Their role is therefore confined to the complex latent representation: they are constrained by the governing equations and interact with the physical prediction through parameters shared by the real and imaginary network paths. No independent physical interpretation is assigned to these imaginary fields.

\subsection{Collocation Sampling}
\label{sec:collocation}

The collocation set
$\mathcal{D}_{\mathrm{eqn}}=\{(x_j,y_j,t_j)\}_{j=1}^{N_c}$
is generated by Latin hypercube sampling~\cite{mckay1979} over the complete space--time domain $\Omega\times[0,T]$, including $\Omega_{\mathrm{hole}}$. Since the gap represents missing measurements rather than an excluded physical region, the flow within it remains governed by Equations~\eqref{eq:ns_x}--\eqref{eq:continuity}. The equation residuals therefore provide the physical constraint where supervised observations are absent~\cite{xu2023practical}. To increase resolution in the most challenging reconstruction region, the domain-wide sample is supplemented with additional collocation points near $\Omega_{\mathrm{hole}}$. Mini-batches from $\mathcal{D}_{\mathrm{eqn}}$ are used during both stages of optimization (Section~\ref{sec:optimization}).

\subsection{Loss Function}
\label{sec:loss}

The training objective consists of a normalized data-misfit term and a physics-residual term:
\begin{align}
  \mathcal{L}_{\mathrm{data}}(\theta) &= \frac{1}{N_d}\sum_{i=1}^{N_d}\left[\frac{\left(u(x_i,y_i,t_i;\theta)-u_i^{*}\right)^{2}}{s_u^{2}} + \frac{\left(v(x_i,y_i,t_i;\theta)-v_i^{*}\right)^{2}}{s_v^{2}}\right], \label{eq:loss_data}\\
  \mathcal{L}_{\mathrm{physics}}(\theta) &= \frac{1}{N_c}\sum_{j=1}^{N_c}\left(|e_{1,j}|^{2}+|e_{2,j}|^{2}+|e_{3,j}|^{2}\right), \label{eq:loss_phys}\\
  \mathcal{L}(\theta) &= \mathcal{L}_{\mathrm{data}}(\theta) + \lambda\,\mathcal{L}_{\mathrm{physics}}(\theta), \label{eq:loss_total}
\end{align}
where $u$ and $v$ are the real parts of the network outputs, $s_u^{2}$ and $s_v^{2}$ denote the sample variances of the supervised values $\{u_i^{*}\}$ and $\{v_i^{*}\}$, and $e_{k,j}$ is the residual $e_k$ evaluated at the $j$th collocation point. The parameter $\lambda>0$ controls the relative contribution of the physics constraint. During mini-batch optimization, the sums in Equations~\eqref{eq:loss_data}--\eqref{eq:loss_phys} are replaced by averages over batches sampled from $\mathcal{D}_{\mathrm{data}}$ and $\mathcal{D}_{\mathrm{eqn}}$.

The variance scaling in Equation~\eqref{eq:loss_data} renders the velocity misfit dimensionless and reduces its dependence on the characteristic magnitude of each component. This is useful across the present test cases, for which the velocity scales differ substantially: the cylinder wake has $u_\infty=1$, whereas the turbulent case in Section~\ref{sec:turbulence} has a root-mean-square velocity of approximately $4.1$. For the cylinder-wake data, the standard deviations of $u$ and $v$ are similar ($0.28$ and $0.27$, respectively; Table~\ref{TabCyl}), so the normalization introduces little relative reweighting between the two components; its effect becomes more relevant when their fluctuation levels differ appreciably.

The data term measures agreement with the available observations, including any imposed measurement noise, whereas $\mathcal{L}_{\mathrm{physics}}$ penalizes predictions that are inconsistent with the governing equations. Their relative influence is controlled by $\lambda$: smaller values emphasize agreement with the observations, while larger values place greater emphasis on physical consistency. The combined objective therefore balances fidelity to the measured data with equation-constrained reconstruction, particularly within $\Omega_{\mathrm{hole}}$, where direct supervision is unavailable.

\subsection{Two-Stage Optimization}
\label{sec:optimization}

The parameters $\theta$, comprising the real and imaginary parts of all complex weights and biases, are optimized in two successive phases.

\textbf{Phase I: first-order pre-training.}
The loss in Equation~\eqref{eq:loss_total} is first minimized for $N_1$ epochs using Adam~\cite{kingma2014adam}, with an optional cosine-annealing learning-rate schedule~\cite{loshchilov2017sgdr}. Mini-batches are sampled independently from $\mathcal{D}_{\mathrm{data}}$ and $\mathcal{D}_{\mathrm{eqn}}$, and gradient-norm clipping is applied for numerical stability. This stage provides a suitable initialization for the subsequent curvature-based refinement.

\textbf{Phase II: self-scaled quasi-Newton refinement.}
The parameters are then refined for $N_2$ outer iterations using either self-scaled Broyden (SSBroyden) or self-scaled BFGS (SSBFGS), following the self-scaled quasi-Newton framework used for PINNs in~\cite{urban2025}. SSBroyden is employed for the Taylor--Green vortex and turbulent-flow cases, whereas SSBFGS is used for the cylinder wake. A limited-memory formulation with Oren--Luenberger self-scaling~\cite{oren1974,oren1976} is adopted. The search direction is
$\mathbf{d}_k=-\mathbf{H}_k\nabla\mathcal{L}(\theta_k)$, where $\mathbf{H}_k$ approximates the inverse Hessian and is evaluated through the standard two-loop recursion~\cite{liu1989lbfgs,nocedal2006} using the $m$ most recent curvature pairs
\begin{equation}
  \mathbf{s}_i=\theta_{i+1}-\theta_i,\qquad
  \mathbf{y}_i=\nabla\mathcal{L}(\theta_{i+1})-\nabla\mathcal{L}(\theta_i).
  \label{eq:curv_pairs}
\end{equation}
Unlike standard L-BFGS, which commonly initializes the inverse-Hessian approximation with
$\gamma^{\mathrm{BB}}_k=
\mathbf{s}_{k-1}^{\top}\mathbf{y}_{k-1}/
\mathbf{y}_{k-1}^{\top}\mathbf{y}_{k-1}$,
the self-scaled formulation uses
\begin{equation}
  \gamma^{\mathrm{OL}}_k =
  \frac{\mathbf{s}_{k-1}^{\top}\mathbf{y}_{k-1}}
       {\mathbf{y}_{k-1}^{\top}\mathbf{H}_{k-1}\mathbf{y}_{k-1}},
  \label{eq:gamma_ol}
\end{equation}
to define $\mathbf{H}_k^{0}=\gamma_k\mathbf{I}$~\cite{oren1974,oren1976}. This scaling adapts the inverse-Hessian approximation to the curvature observed during optimization.

The step length is determined using a strong-Wolfe line search. The mini-batch is kept fixed throughout each outer iteration so that all line-search evaluations correspond to the same objective, and curvature information is retained across batch updates following stochastic limited-memory quasi-Newton strategies~\cite{schraudolph2007,mokhtari2015}. For the cylinder wake, all supervised observations and a sampled subset of collocation points are used, with the collocation batch refreshed as specified in Section~\ref{sec:cyl_use}.

Numerical safeguards are applied during Phase II: non-finite line-search evaluations are rejected, excessively large search directions are norm-limited, and the previous checkpoint is restored with the curvature history reset if an update produces a non-finite loss. For the cylinder wake, optimization is terminated when the loss stagnates, and the parameters corresponding to the minimum loss are retained. The values of $N_1$, $N_2$, memory size, batch sizes, and other case-specific settings are reported in Section~\ref{sec:numerical_study}.

\subsection{Evaluation Metrics}
\label{sec:metrics}

After training, the reconstructed fields are evaluated on the complete reference grid, including $\Omega_{\mathrm{hole}}$, at each snapshot over $[0,T]$. For a scalar field $q\in\{u,v,p,\omega\}$, let $q_n$ and $\tilde{q}_n$, $n=1,\dots,N$, denote the reference and reconstructed values at the $N$ grid points, respectively. Their spatial means are denoted by $\bar{q}$ and $\bar{\tilde{q}}$, and the reference dynamic range is $Q=\max_n q_n-\min_n q_n$. Following commonly used measures in PINN-based flow reconstruction~\cite{raissi2019,xu2023practical,liu2025ffpicn}, we compute the relative $L_2$ error, coefficient of determination, and maximum pointwise error as
\begin{equation}
  \text{Relative }L_2\text{ error}(q)=\frac{\Big[\sum_{n=1}^{N}(\tilde{q}_n-q_n)^{2}\Big]^{1/2}}{\Big[\sum_{n=1}^{N}q_n^{2}\Big]^{1/2}},\qquad
  R^{2}(q)=1-\frac{\sum_{n=1}^{N}(\tilde{q}_n-q_n)^{2}}{\sum_{n=1}^{N}(q_n-\bar{q})^{2}},\qquad
  L_\infty(q)=\max_{n}|\tilde{q}_n-q_n|,
  \label{eq:metrics1}
\end{equation}
together with the root-mean-square error, mean absolute error, and Pearson correlation coefficient,
\begin{equation}
  \mathrm{RMSE}(q)=\Big[\frac{1}{N}\sum_{n=1}^{N}(\tilde{q}_n-q_n)^{2}\Big]^{1/2},\qquad
  \mathrm{MAE}(q)=\frac{1}{N}\sum_{n=1}^{N}|\tilde{q}_n-q_n|,\qquad
  \rho(q)=\frac{\sum_{n}(q_n-\bar{q})(\tilde{q}_n-\bar{\tilde{q}})}{\big[\sum_{n}(q_n-\bar{q})^{2}\big]^{1/2}\big[\sum_{n}(\tilde{q}_n-\bar{\tilde{q}})^{2}\big]^{1/2}}.
  \label{eq:metrics2}
\end{equation}
The peak signal-to-noise ratio (PSNR) and structural similarity index (SSIM)~\cite{wang2004ssim} are additionally used to assess field-level reconstruction quality:
\begin{equation}
  \mathrm{PSNR}(q)=10\log_{10}\frac{Q^{2}}{\mathrm{RMSE}^{2}(q)},\qquad
  \mathrm{SSIM}(q)=\frac{1}{M}\sum_{m=1}^{M}\frac{\left(2\mu_m\tilde{\mu}_m+C_1\right)\left(2\sigma_{m}^{q\tilde{q}}+C_2\right)}{\left(\mu_m^{2}+\tilde{\mu}_m^{2}+C_1\right)\left(\sigma_m^{2}+\tilde{\sigma}_m^{2}+C_2\right)}.
  \label{eq:metrics3}
\end{equation}
Here, SSIM is evaluated over $M$ local windows, where $\mu_m$ and $\tilde{\mu}_m$ are the local means, $\sigma_m^{2}$ and $\tilde{\sigma}_m^{2}$ are the local variances, and $\sigma_m^{q\tilde{q}}$ is the local covariance. The stabilization constants are $C_1=(0.01\,Q)^{2}$ and $C_2=(0.03\,Q)^{2}$, following Ref.~\cite{wang2004ssim}. All metrics are evaluated against the noise-free reference for each field and snapshot, and temporal averages are denoted by $\langle\cdot\rangle$. Pointwise error maps are reported as $|\Delta q|=|\tilde{q}-q|$, e.g., $|\Delta u|$ and $|\Delta\omega|$.

These metrics provide complementary measures of reconstruction quality. Relative $L_2$, RMSE, MAE, and $L_\infty$ quantify error magnitude; $R^{2}$ measures agreement with the spatial variance of the reference field; PSNR relates reconstruction error to its dynamic range; and SSIM assesses the preservation of local field structure. The correlation coefficient $\rho$ measures spatial association independently of a uniform offset or positive scaling, helping distinguish structural agreement from amplitude accuracy.

\subsection{Summary of the Design and Training Procedure}
\label{sec:summary}

Table~\ref{Tab1} summarizes the role of each FCV-PICRN component. The Fourier embedding supplies frequency-rich coordinate features, while the complex-valued layers enable coupled amplitude--phase transformations. Complex convolutional blocks provide a compact parameterization, and residual connections facilitate information propagation through the stacked architecture. The physics residual constrains the predicted fields throughout the domain, including $\Omega_{\mathrm{hole}}$, where supervised measurements are unavailable, and provides an explicit physical constraint against noise-corrupted observations. The complete training and evaluation workflow is summarized in Algorithm~\ref{alg:fcvpicrn}.

\begin{table}[pos=htbp]
\centering
\caption{Role of the main components of FCV-PICRN in the gappy, noisy reconstruction problem.}
\label{Tab1}
\small
\begin{tabular}{>{\raggedright\arraybackslash}p{0.27\textwidth}>{\raggedright\arraybackslash}p{0.31\textwidth}>{\raggedright\arraybackslash}p{0.34\textwidth}}
\hline
Component & Difficulty addressed & Mechanism \\
\hline
Fourier embedding, Eq.~\eqref{eq:fourier} & Spectral bias; oscillatory dynamics & Fixed sinusoidal features (periodic prior) \\
Complex-valued layers, Eq.~\eqref{eq:complex_product} & Amplitude--phase coupling & Complex weights scale and rotate features \\
Complex 1-D convolutions, Eq.~\eqref{eq:cconv} & Over-parameterization & Local, weight-shared feature mixing \\
Residual connections, Eq.~\eqref{eq:res_block} & Gradient degradation with depth & Identity paths \\
CGELU activation, Eq.~\eqref{eq:cgelu} & Vanishing second derivatives; overflow & Smooth, at most linear growth \\
Navier--Stokes residuals, Eqs.~\eqref{eq:e1}--\eqref{eq:e3} & Missing data in $\Omega_{\mathrm{hole}}$; noise; unmeasured $p$ & Real part enforces the Navier--Stokes equations \\
Normalized data loss, Eq.~\eqref{eq:loss_data} & Case-dependent velocity scale & Dimensionless, balanced misfit \\
Adam + SSBroyden/SSBFGS, Eq.~\eqref{eq:gamma_ol} & Stiff, ill-conditioned objective & First-order start, quasi-Newton refinement \\
\hline
\end{tabular}
\end{table}

\begin{algorithm}[!htbp]
\caption{Training and evaluation procedure for FCV-PICRN}
\label{alg:fcvpicrn}
\small
\begin{algorithmic}[1]

\REQUIRE Governing equations and $Re$; domain $\Omega\times[0,T]$; supervised dataset $\mathcal{D}_{\mathrm{data}}$ (sparse, gappy, noisy, or possibly empty); optional initial-, boundary-, or anchor-point data $\mathcal{D}_{\mathrm{aux}}$; gap region $\Omega_{\mathrm{hole}}$ (possibly empty); hyperparameters $K$, $W$, $L$, $k$, $\lambda$, $N_1$, $N_2$, and learning-rate schedule

\ENSURE Optimized parameters $\theta^{*}$; reconstructed fields
$u=\ReP(\hat{u})$, $v=\ReP(\hat{v})$, $p=\ReP(\hat{p})$, and
$\omega=v_x-u_y$; evaluation metrics

\STATE Normalize the input coordinates using Equation~\eqref{eq:minmax}; retain the scaling factors when data standardization is applied

\STATE Generate $\mathcal{D}_{\mathrm{eqn}}$ by Latin hypercube sampling over $\Omega\times[0,T]$, including $\Omega_{\mathrm{hole}}$, and enrich the sampling near the gap

\STATE Initialize the complex-valued weights in polar form as described in Section~\ref{sec:complex}

\STATE \textbf{Phase I: Adam pre-training}

\FOR{epoch $=1$ \TO $N_1$}
    \FOR{each mini-batch from $\mathcal{D}_{\mathrm{data}}$, $\mathcal{D}_{\mathrm{aux}}$, and $\mathcal{D}_{\mathrm{eqn}}$}
        \STATE Apply the Fourier embedding and complex residual network to obtain $(\hat{u},\hat{v},\hat{p})$ using Equations~\eqref{eq:minmax}--\eqref{eq:res_out}; extract the physical fields from Equation~\eqref{eq:real_parts}
        
        \STATE Evaluate $e_1$, $e_2$, and $e_3$ by automatic differentiation using Equations~\eqref{eq:e1}--\eqref{eq:e3}
        
        \STATE Compute $\mathcal{L}=\mathcal{L}_{\mathrm{data}}+\lambda\mathcal{L}_{\mathrm{physics}}$ from Equations~\eqref{eq:loss_data}--\eqref{eq:loss_total}, including auxiliary loss terms when applicable
        
        \STATE Back-propagate $\mathcal{L}$, clip $\|\nabla_{\theta}\mathcal{L}\|$, update the learning rate, and perform an Adam step
    \ENDFOR
\ENDFOR

\STATE \textbf{Phase II: self-scaled quasi-Newton refinement}

\STATE Initialize the quasi-Newton curvature history

\FOR{epoch $=1$ \TO $N_2$}
    \STATE Refresh the collocation mini-batch when prescribed, keep it fixed during the current line search, and checkpoint $\theta$
    
    \STATE Compute the self-scaled quasi-Newton direction using the two-loop recursion and Equation~\eqref{eq:gamma_ol}; restrict its norm when required
    
    \STATE Determine the step length using a strong-Wolfe line search and reject non-finite trial evaluations
    
    \IF{$\mathcal{L}$ is non-finite}
        \STATE Restore the checkpointed parameters and reset the curvature history
    \ELSE
        \STATE Accept the update and retain the curvature information
    \ENDIF
    
    \IF{the prescribed early-stopping criterion is satisfied}
        \STATE Stop the refinement and retain the parameters corresponding to the minimum loss
    \ENDIF
\ENDFOR

\STATE Set the retained parameters as $\theta^{*}$

\STATE Evaluate $\mathcal{F}_{\theta^{*}}$ over the complete evaluation grid, including $\Omega_{\mathrm{hole}}$

\STATE Compute the reconstruction metrics using Equations~\eqref{eq:metrics1}--\eqref{eq:metrics3} with respect to the reference fields

\end{algorithmic}
\end{algorithm}

\section{Numerical Validation}
\label{sec:numerical_study}
The proposed FCV-PICRN is evaluated on three test cases of increasing complexity. The unsteady Taylor--Green vortex is considered first (Section~\ref{sec:tgv}), where its exact solution provides a controlled benchmark for validating the architecture and optimization strategy without uncertainty in the reference data. The framework is then applied to cylinder-wake reconstruction at $Re=100$ (Section~\ref{sec:flow_reconstruction}), progressing from extremely sparse supervision to boundary-adjacent data loss and highly noisy measurements, and finally to gaps within the dynamically critical wake combined with Gaussian and speckle noise. A consistent architecture, activation, initialization, and two-stage optimization strategy is retained throughout, with $K=4$ Fourier frequencies for the laminar cases; only the network size and training parameters are adjusted to each problem. The nearly data-free Taylor--Green vortex is trained for 30,000 Adam epochs~\cite{kingma2014adam} followed by 10,000 SSBroyden iterations~\cite{urban2025} (Section~\ref{sec:tgv_training}), whereas the cylinder-wake cases use 500 Adam epochs followed by at most 500 SSBFGS iterations (Section~\ref{sec:cyl_use}); the turbulent-flow settings are summarized in Table~\ref{Tab6}. In all cases, collocation points span the complete space--time domain, including the prescribed gaps (Section~\ref{sec:collocation}), allowing the governing equations to constrain regions without supervised data. Within the cylinder-wake study, only the sampling fraction, gap geometry, and noise model are varied across the reconstruction cases.

\subsection{Taylor--Green Vortex}
\label{sec:tgv}

\subsubsection{Problem Statement}
As an independent verification before addressing the reconstruction cases, FCV-PICRN is tested on the two-dimensional Taylor--Green vortex (TGV) forward problem~\cite{taylor1937,brachet1983}. The model is provided only with the governing equations, Equations~\eqref{eq:ns_x}--\eqref{eq:continuity}, the initial condition at $t=0$, and the space--time domain
\begin{equation}
    \Omega=[0,2\pi]\times[0,2\pi],\qquad t\in[0,T].
    \label{eq:tgv_domain}
\end{equation}
No velocity observations are prescribed for $t>0$; hence, the evolution of $u$, $v$, $p$, and $\omega$ is determined primarily through the Navier--Stokes constraints. This benchmark simultaneously tests nonlinear advection, viscous diffusion, and incompressibility. Moreover, the availability of an exact divergence-free solution permits direct verification of the predicted fields without discretization error in the reference solution.

\subsubsection{Analytical Solution}
For $Re=1/\nu=100$, the analytical solution is given by
\begin{align}
  u(x,y,t) &= -\cos x\,\sin y\;e^{-2\nu t}, \label{eq:tgv_u}\\
  v(x,y,t) &= \sin x\,\cos y\;e^{-2\nu t}, \label{eq:tgv_v}\\
  p(x,y,t) &= -\tfrac{1}{4}\left(\cos 2x+\cos 2y\right)e^{-4\nu t}. \label{eq:tgv_p}
\end{align}
To obtain this solution, we consider the separable form
$u=-\cos x\sin y\,F(t)$, $v=\sin x\cos y\,F(t)$, and
$p=-\tfrac14(\cos2x+\cos2y)\,G(t)$. This representation satisfies the incompressibility condition $u_x+v_y\equiv0$ for arbitrary $F(t)$. Furthermore, $u_{xx}+u_{yy}=-2u$ and $uu_x+vu_y=-\sin x\cos x\,F^{2}$. Substitution into Equation~\eqref{eq:ns_x} therefore gives
\begin{equation}
  \cos x\sin y\left(-F'-2\nu F\right)+\sin x\cos x\left(G-F^{2}\right)=0.
  \label{eq:tgv_reduced}
\end{equation}
Because the two spatial terms $\cos x\sin y$ and $\sin x\cos x$ are linearly independent, their coefficients vanish separately, yielding $F'=-2\nu F$ and $G=F^{2}$. With $F(0)=1$, these relations give $F=e^{-2\nu t}$ and $G=e^{-4\nu t}$, while Equation~\eqref{eq:ns_y} leads to the same conditions. The corresponding vorticity follows directly from the velocity gradients as
\begin{equation}
  \omega=v_x-u_y=2\cos x\cos y\;e^{-2\nu t}.
  \label{eq:tgv_w}
\end{equation}

\begin{figure}[pos=!htbp]
\refstepcounter{figure}
\begin{center}
    \includegraphics[max width=\textwidth, max height=0.82\textheight]{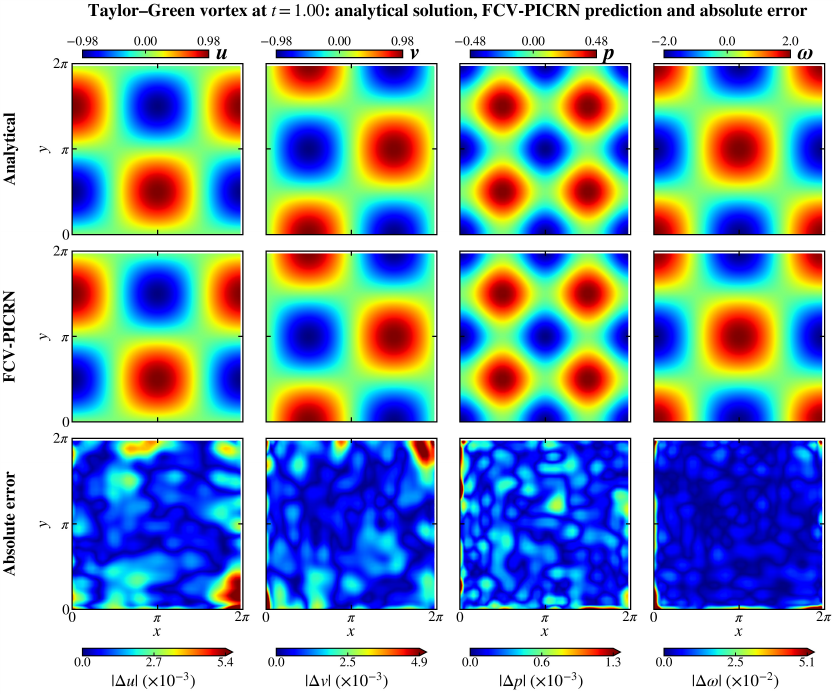}
\end{center}
\begin{quote}
\noindent\textbf{Figure \thefigure.}
Comparison of the analytical and FCV-PICRN-predicted Taylor--Green vortex fields at $t=1.00$, with the corresponding absolute errors for $u$, $v$, $p$, and $\omega$.
\end{quote}
\label{fig4}
\end{figure}

\subsubsection{Training Protocol}
\label{sec:tgv_training}

The network is configured with $K=4$, hidden width $W=64$, and $L=4$ single-channel blocks with $k=3$, resulting in 4,006 real-valued parameters. Training is performed for 30,000 Adam epochs using a cosine-annealing learning-rate schedule~\cite{loshchilov2017sgdr}, followed by 10,000 SSBroyden epochs~\cite{urban2025}. This extended optimization is required because no interior observations are available and the solution is therefore governed primarily by the physics residual. The objective function consists of the physics loss in Equation~\eqref{eq:loss_phys} together with the initial-condition mismatch at $t=0$. A small number of pressure-anchor points, sampled throughout the temporal domain, are additionally supervised to eliminate the additive-constant indeterminacy of pressure (Section~\ref{sec:problem}), since only its gradient appears in the governing equations. The collocation, initial-condition, and pressure-anchor points are resampled at every iteration following Algorithm~\ref{alg:fcvpicrn}. Periodicity is not enforced explicitly as a hard constraint.

\subsubsection{Results and Discussion}
Figure~\ref{fig4} compares the analytical and predicted fields at $t=1.00$, while Table~\ref{Tab3} summarizes the corresponding error metrics. Excellent agreement is obtained for all four variables, with relative $L_2$ errors of $3.03\times10^{-3}$ and $2.45\times10^{-3}$ for $u$ and $v$, $1.25\times10^{-3}$ for $p$, and $8.65\times10^{-3}$ for $\omega$. The corresponding $R^{2}$ values exceed $0.9999$, SSIM remains above $0.999$, and PSNR ranges from $53$ to $70$~dB. The comparatively larger error in $\omega$ is associated with its dependence on velocity derivatives, which amplifies small prediction errors. In contrast, the low pressure error indicates that the temporal pressure anchors effectively remove the gauge ambiguity.

The absolute-error maps in the bottom row of Figure~\ref{fig4} indicate that the largest deviations, reflected by the $L_\infty$ values in Table~\ref{Tab3}, occur near the domain edges and corners. Interior errors are approximately two to eight times smaller; for example, the maximum interior error in $\omega$ is about $0.018$, compared with $0.14$ near a corner. This localization is consistent with periodicity not being imposed explicitly, leaving the solution near $\partial\Omega$ constrained only by the governing residual and initial condition. Such boundary errors could be reduced by directly embedding periodicity, for example through Fourier features based on $\sin x$ and $\cos x$. Overall, the TGV benchmark verifies the accuracy of the proposed architecture and two-stage optimization strategy for the incompressible Navier--Stokes equations before their application to the subsequent reconstruction problems.

\begin{table}[pos=htbp]
\centering
\caption{Error metrics of the FCV-PICRN prediction for the Taylor--Green vortex at $t=1.00$.}
\label{Tab3}
\begin{tabular}{lccccc}
\hline
Field & Relative $L_2$ error & $L_\infty$ & $R^{2}$ & SSIM & PSNR (dB) \\
\hline
$u$ & $3.03\times10^{-3}$ & $1.09\times10^{-2}$ & 0.999991 & 0.99988 & 62.4 \\
$v$ & $2.45\times10^{-3}$ & $1.03\times10^{-2}$ & 0.999994 & 0.99993 & 64.2 \\
$p$ & $1.25\times10^{-3}$ & $2.06\times10^{-3}$ & 0.999998 & 0.99998 & 70.1 \\
$\omega$ & $8.65\times10^{-3}$ & $1.42\times10^{-1}$ & 0.999925 & 0.99930 & 53.3 \\
\hline
\end{tabular}
\end{table}

\subsection{Flow Reconstruction Past a Circular Cylinder at $Re=100$}
\label{sec:flow_reconstruction}

\subsubsection{Data Description}
\label{sec:cyl_data}
The second test case considers flow past a circular cylinder at $Re=u_\infty D/\nu=100$, where alternating shear-layer roll-up produces a periodic K\'arm\'an vortex street~\cite{williamson1996}. This benchmark is widely used for physics-informed flow reconstruction because it combines convecting vortices, thin shear layers, and an unobserved pressure field. The free-stream velocity, cylinder diameter, and kinematic viscosity are $u_\infty=1$, $D=1$, and $\nu=0.01$, respectively. Length, velocity, time, and kinematic pressure are nondimensionalized by $D$, $u_\infty$, $D/u_\infty$, and $u_\infty^{2}$, respectively. The resulting flow therefore satisfies Equations~\eqref{eq:ns_x}--\eqref{eq:continuity} at $Re=100$, and all times reported below are non-dimensional.

\begin{figure}[pos=!htbp]
\refstepcounter{figure}
\begin{center}
    \includegraphics[max width=\textwidth, max height=0.82\textheight]{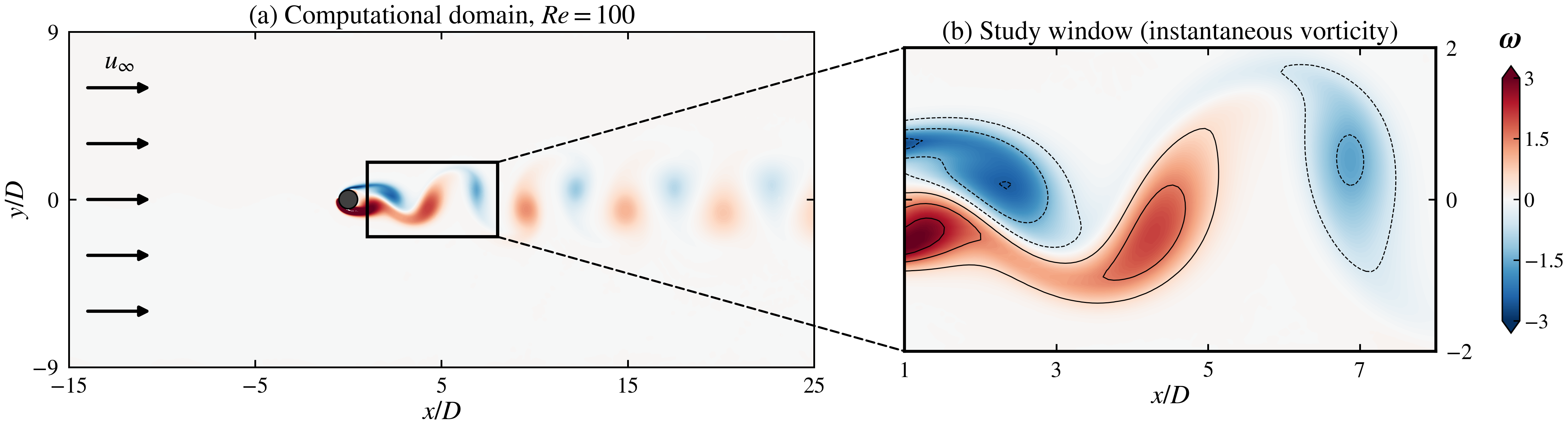}
\end{center}
\begin{quote}
\noindent\textbf{Figure \thefigure.}
Vorticity field for flow past a circular cylinder at $Re=100$: (a) full computational domain and (b) reconstruction window $\Omega=[1,8]\times[-2,2]$. Solid and dashed contours denote positive and negative vorticity, respectively.
\end{quote}
\label{figCyl}
\end{figure}

The reference data are obtained from the direct numerical simulation of Raissi \textit{et al.}~\cite{raissi2019}, generated using the spectral/$hp$ element method~\cite{karniadakis2005}. The computational domain extends $15D$ upstream and $25D$ downstream from the cylinder center (Figure~\ref{figCyl}a) and contains 412 triangular elements with tenth-order polynomial expansions. A uniform free stream is imposed at the inlet, zero pressure at the outlet, and periodic conditions along the lateral boundaries. Time integration employs a third-order stiffly stable scheme until periodic vortex shedding is established~\cite{raissi2019}. The instantaneous vorticity field in Figure~\ref{figCyl}a is taken from the corresponding spectral-element data set, whose 412-element, tenth-order discretization agrees with Ref.~\cite{raissi2019}. The same data set has also been employed by Liu \textit{et al.}~\cite{liu2025ffpicn} and Xu \textit{et al.}~\cite{xu2023practical}, enabling direct comparison with these reconstruction studies.

Following Refs.~\cite{raissi2019,xu2023practical,liu2025ffpicn}, the analysis is restricted to the wake region $\Omega=[1,8]\times[-2,2]$ (Figure~\ref{figCyl}b). Beginning half a cylinder diameter downstream of the rear surface, this window contains the vortex-formation region and the first two pairs of shed vortices, thereby covering the dynamically active near wake where the gaps considered in Sections~\ref{sec:critical_gaps} and \ref{sec:plus_noise} are introduced. Within $\Omega$, the data set provides $u$, $v$, and $p$ on a uniform $100\times50$ grid at 200 time instants, with $t\in[0,19.9]$ and $\Delta t=0.1$, corresponding to $5\,000$ spatial locations and $10^{6}$ space--time samples for each variable. These data-set characteristics were verified directly from the supplied data file and are summarized in Table~\ref{TabCyl}; the spatial and temporal resolutions are consistent with Liu \textit{et al.}~\cite{liu2025ffpicn}.

\begin{table}[pos=htbp]
\centering
\caption{Spatiotemporal specifications of the circular-cylinder wake data set at $Re=100$ within the reconstruction window.}
\label{TabCyl}
\small
\begin{tabular}{lc}
\hline
Property & Value \\
\hline
Reynolds number $Re=u_\infty D/\nu$ & 100 \\
Study window $\Omega$ & $[1,8]\times[-2,2]$ \\
Spatial grid ($N_x\times N_y$) & $100\times50$ \\
Grid spacing ($\Delta x$, $\Delta y$) & $0.0707$, $0.0816$ \\
Snapshots, time step & 200, $\Delta t=0.1$ ($0\le t\le19.9$) \\
Space--time samples per variable & $10^{6}$ \\
Range of $u$ / $v$ & $[-0.24,\,1.32]$ / $[-0.62,\,0.62]$ \\
Standard deviation of $u$ / $v$ & $0.28$ / $0.27$ \\
Range of $p$ & $[-0.54,\,0.07]$ \\
Shedding period $T_s$ / Strouhal number $St$ & $5.90$ / $0.169$ \\
\hline
\end{tabular}
\end{table}

\subsubsection{Data Preparation for FCV-PICRN Reconstruction at $Re=100$}
\label{sec:cyl_use}
FCV-PICRN takes the coordinates $(x,y,t)$ over $\Omega\times[0,19.9]$, normalized using Equation~\eqref{eq:minmax}. For each case, the supervised set $\mathcal{D}_{\mathrm{data}}$ in Equation~\eqref{eq:data_set} is selected from the $10^{6}$ velocity samples according to the prescribed sampling fraction, gap geometry (Section~\ref{sec:gaps}), and noise model (Section~\ref{sec:noise}). Except for the sparse-data case, supervision is provided at 10 uniformly spaced snapshots ($5\%$ of the temporal record), after removing all samples within $\Omega_{\mathrm{hole}}$. Only $u$ and $v$ are used for supervision; $p$ and $\omega$ are inferred from the trained model through Equation~\eqref{eq:real_parts} and $\omega=v_x-u_y$, respectively. The clean velocity, pressure, and vorticity fields are used only as references for the metrics defined in Section~\ref{sec:metrics}. Reconstruction is assessed over all $10^{6}$ grid points, including $\Omega_{\mathrm{hole}}$, with the additive pressure offset removed before comparison (Section~\ref{sec:problem}).

Collocation points are sampled throughout $\Omega\times[0,19.9]$, including $\Omega_{\mathrm{hole}}$ (Section~\ref{sec:collocation}), with an additional $25\%$ concentrated around the gap. All reconstruction cases use the same network and training settings listed in Table~\ref{Tab2}, such that only the data configuration varies between cases. Adam is applied with 50 mini-batches per epoch, cosine learning-rate decay~\cite{loshchilov2017sgdr}, and gradient clipping at norm 5. SSBFGS uses 20 iterations per epoch with a strong-Wolfe line search and a step bound $\|\mathbf{d}\|\le5$, using mini-batches containing all supervised data and 40,000 collocation points, which are resampled every 500 iterations. The parameters corresponding to the minimum loss are retained. Each case is trained for 500 Adam epochs followed by at most 500 SSBFGS epochs on two NVIDIA T4 GPUs; for reference, FFPICN was trained for 3,000 epochs on the same data set~\cite{liu2025ffpicn}.

\begin{table}[pos=htbp]
\centering
\caption{FCV-PICRN hyperparameters for flow past a circular cylinder at $Re=100$, identical across all reconstruction cases.}
\label{Tab2}
\small
\begin{tabular}{ll}
\hline
Parameter & Value \\
\hline
Fourier order $K$ (frequencies $w_k$) & 4 (1, 2, 4, 8) \\
Hidden width $W$ / residual blocks $L$ & 160 / 4 \\
Activation / initialization & CGELU / complex Glorot~\cite{trabelsi2018} \\
Trainable parameters & 216,006 \\
Supervised snapshots & 10 of 200 ($5\%$) \\
Collocation points $N_c$ & 250,000 (LHS) \\
Physics weight $\lambda$ & 3 \\
Phase I: Adam~\cite{kingma2014adam} & 500 epochs, lr $10^{-3}\rightarrow10^{-5}$ \\
Phase II: SSBFGS~\cite{urban2025} & $\le500$ epochs, memory $m=50$ \\
Stopping criterion & $<0.2\%$ loss decrease over 30 epochs \\
Hardware & 2$\times$NVIDIA T4, single precision \\
\hline
\end{tabular}
\end{table}

\subsubsection{Cylinder-Wake Reconstruction at $Re=100$ from $0.16\%$ Sparse Velocity Data}
\label{sec:sparse}

\begin{figure}[pos=!htbp]
\refstepcounter{figure}
\begin{center}
    \includegraphics[max width=\textwidth, max height=0.82\textheight]{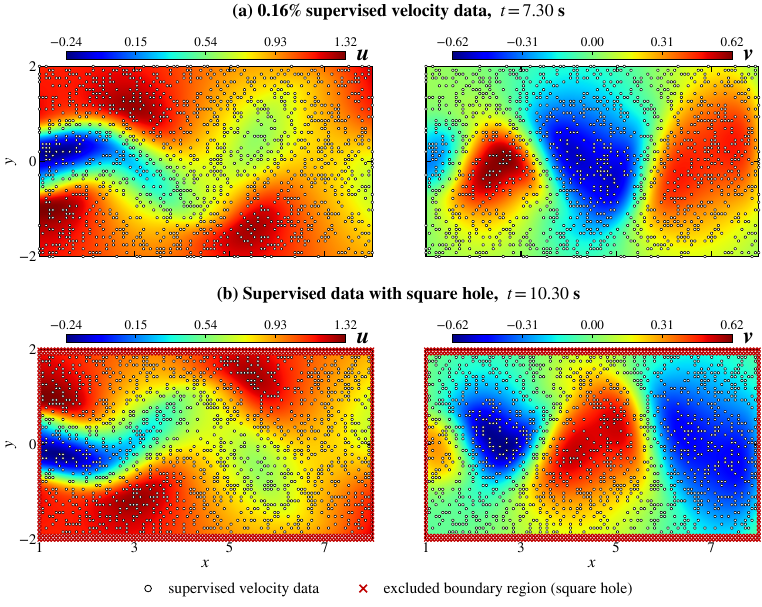}
\end{center}
\begin{quote}
\noindent\textbf{Figure \thefigure.}
Supervised velocity data for flow past a circular cylinder at $Re=100$: (a) $0.16\%$ sparse-data case at $t=7.30$ and (b) square-hole case at $t=10.30$. Circles and crosses denote supervised and excluded points, respectively.
\end{quote}
\label{fig5}
\end{figure}

\begin{figure}[pos=!htbp]
\refstepcounter{figure}
\begin{center}
    \includegraphics[max width=\textwidth, max height=0.82\textheight]{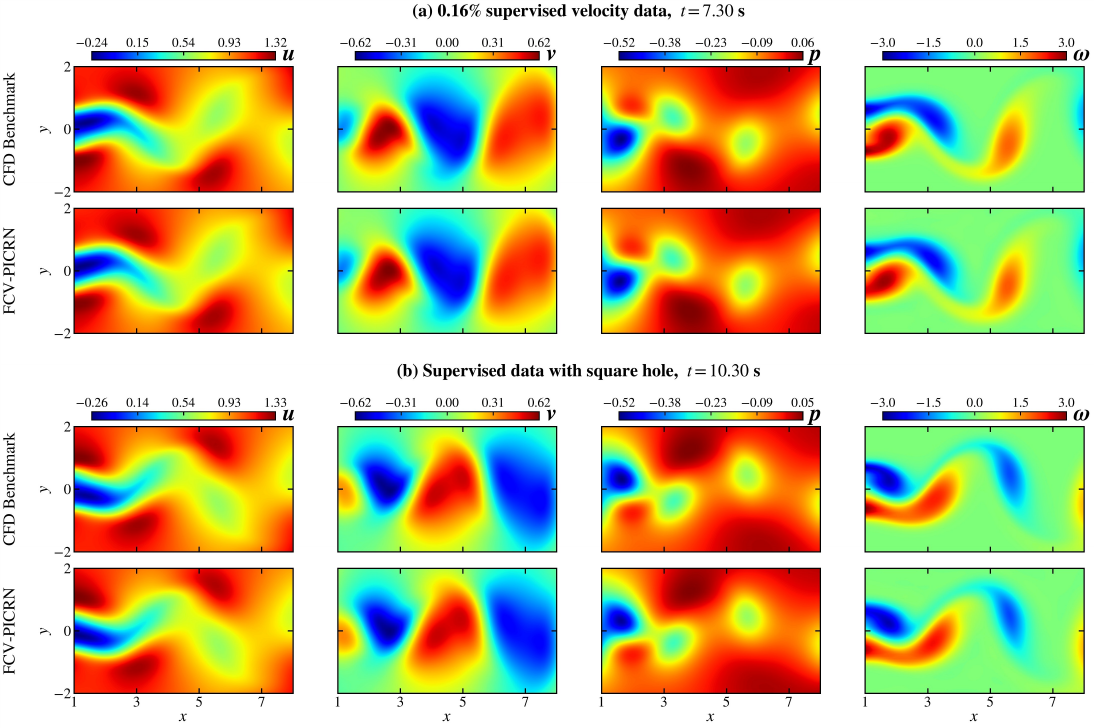}
\end{center}

\begin{quote}
\noindent\textbf{Figure \thefigure.}
Comparison of CFD Benchmark and FCV-PICRN reconstructed velocity fields ($u$ and $v$) for flow past a circular cylinder at $Re=100$: (a) $0.16\%$ sparse-data case at $t=7.30$ and (b) square-hole case at $t=10.30$.
\end{quote}
\label{fig6}
\end{figure}

The first case examines reconstruction under extremely sparse supervision, using only $0.16\%$ of the reference velocity data without gaps or noise (Figure~\ref{fig5}a). Figure~\ref{fig6}a compares the predicted $u$, $v$, $p$, and $\omega$ with the CFD benchmark at $t=7.30$. FCV-PICRN reproduces the principal wake structures, including the alternating vortices in $\omega$ and the associated low-pressure regions in $p$, although neither pressure nor vorticity is supervised. At the same $0.16\%$ data fraction, FFPICN reported $R^{2}$ values of $0.54$, $0.69$, and $0.51$ for $u$, $v$, and $p$, respectively, with the reduced accuracy attributed to insufficient data coverage~\cite{liu2025ffpicn}. As summarized in Table~\ref{Tab4}, FCV-PICRN reduces the corresponding relative $L_2$ errors from $0.2492$ to $0.0423$ for $u$, $0.7491$ to $0.0790$ for $v$, and $0.7276$ to $0.2434$ for $p$, while increasing $R^{2}$ to $0.993$, $0.993$, and $0.943$, respectively. Among the reconstructed variables, pressure exhibits the largest relative error because it is not directly supervised and is constrained through its gradient in the momentum equations.

\subsubsection{Cylinder-Wake Reconstruction at $Re=100$ from Square-Boundary Gappy Velocity Data}
\label{sec:square_hole}

In the second case, supervised velocity measurements are removed from a band along the boundary of the reconstruction window, producing a square-frame gap around the retained interior data (Figure~\ref{fig5}b). Consequently, the flow near the inlet, outlet, and lateral edges of the window is reconstructed from the available interior measurements together with the governing equations. Figure~\ref{fig6}b shows that, at $t=10.30$, FCV-PICRN reproduces the benchmark fields up to the window boundaries, including vortical structures entering and leaving the reconstruction region. The inferred $p$ and $\omega$ fields also remain consistent with the reference solution without noticeable edge distortions. As reported in Table~\ref{Tab4}, FCV-PICRN yields relative $L_2$ errors of $0.0024$, $0.0128$, and $0.0125$ for $u$, $v$, and $p$, respectively, compared with $0.0041$, $0.0168$, and $0.0254$ reported by FFPICN for its square-hole case~\cite{liu2025ffpicn}; the corresponding $R^{2}$ values for both methods range from $0.9991$ to $0.9996$. These configurations are complementary rather than identical: FFPICN retains boundary measurements while removing interior data, whereas the present case retains interior measurements and removes the boundary data. The results therefore demonstrate that FCV-PICRN maintains comparable or lower reconstruction errors when the flow along the window boundaries is inferred rather than directly supervised.

\begin{table}[pos=htbp]
\centering
\caption{Comparison of FFPICN~\cite{liu2025ffpicn} and the proposed FCV-PICRN for flow past a circular cylinder at $Re=100$: relative $L_2$ error and $R^2$ for sparse-data and boundary-gap cases.}
\label{Tab4}
\small\setlength{\tabcolsep}{4pt}
\begin{tabular}{llcccccc}
\hline
 & & \multicolumn{3}{c}{Relative $L_2$ error} & \multicolumn{3}{c}{$R^{2}$} \\
Method & Data & $u$ & $v$ & $p$ & $u$ & $v$ & $p$ \\
\hline
FFPICN~\cite{liu2025ffpicn} & $0.16\%$ & 0.2492 & 0.7491 & 0.7276 & 0.5403 & 0.6947 & 0.5068 \\
\textbf{FCV-PICRN} & $0.16\%$ & \textbf{0.0423} & \textbf{0.0790} & \textbf{0.2434} & \textbf{0.9934} & \textbf{0.9932} & \textbf{0.9434} \\
FFPICN~\cite{liu2025ffpicn} & Square hole$^{a}$ & 0.0041 & 0.0168 & 0.0254 & 0.9994 & 0.9996 & 0.9992 \\
\textbf{FCV-PICRN} & Square hole$^{b}$ & \textbf{0.0024} & \textbf{0.0128} & \textbf{0.0125} & \textbf{0.9998} & \textbf{0.9996} & \textbf{0.9994} \\
\hline
\end{tabular}

\vspace{2pt}{\footnotesize $^{a}$Only the boundary data are retained~\cite{liu2025ffpicn}. $^{b}$The boundary data are removed and the interior data are retained (Figure~\ref{fig5}b).}
\end{table}

\subsubsection{Cylinder-Wake Reconstruction at $Re=100$ under $15\%$ and $20\%$ Gaussian Noise}
\label{sec:high_noise}
\begin{figure}[pos=!htbp]
\refstepcounter{figure}
\begin{center}
    \includegraphics[max width=\textwidth, max height=0.82\textheight]{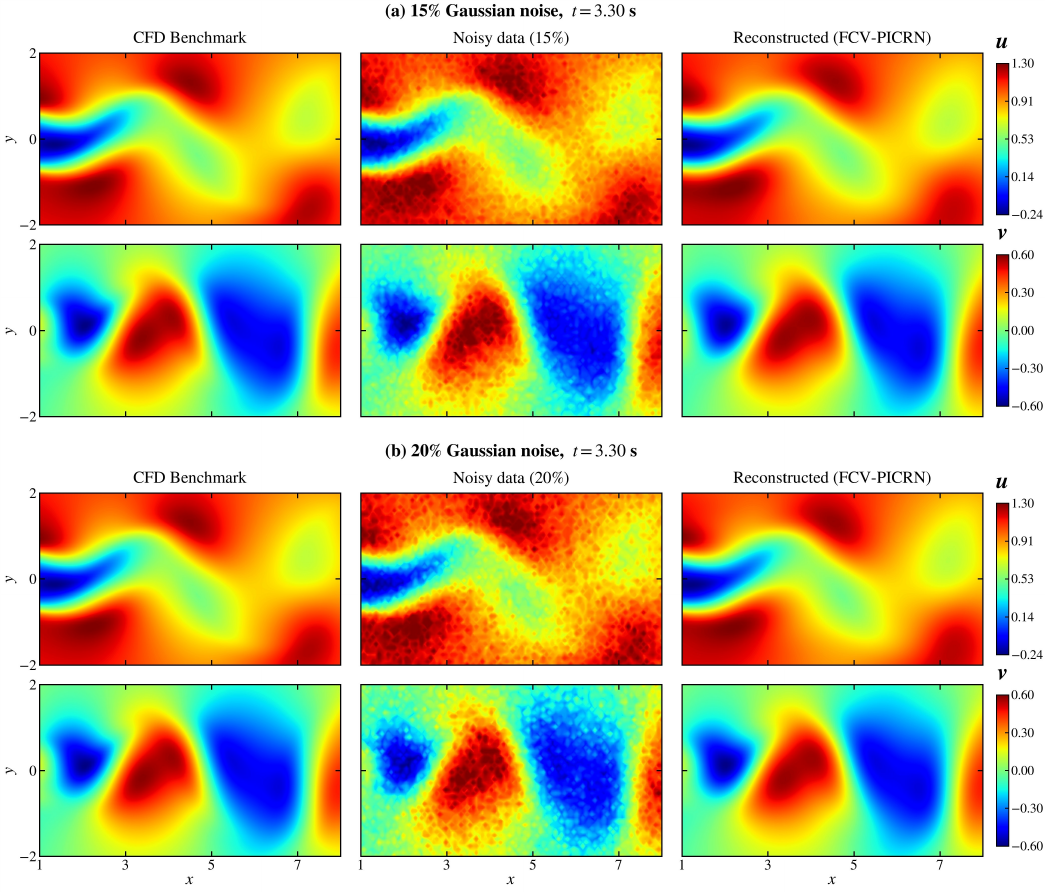}
\end{center}

\begin{quote}
\noindent\textbf{Figure \thefigure.}
Comparison of CFD Benchmark, noisy data, and FCV-PICRN reconstructed velocity fields ($u$ and $v$) for flow past a circular cylinder at $Re=100$ and $t=3.30$: (a) $15\%$ and (b) $20\%$ Gaussian noise.
\end{quote}
\label{fig7}
\end{figure}

\begin{figure}[pos=!htbp]
\refstepcounter{figure}
\begin{center}
    \includegraphics[max width=\textwidth, max height=0.82\textheight]{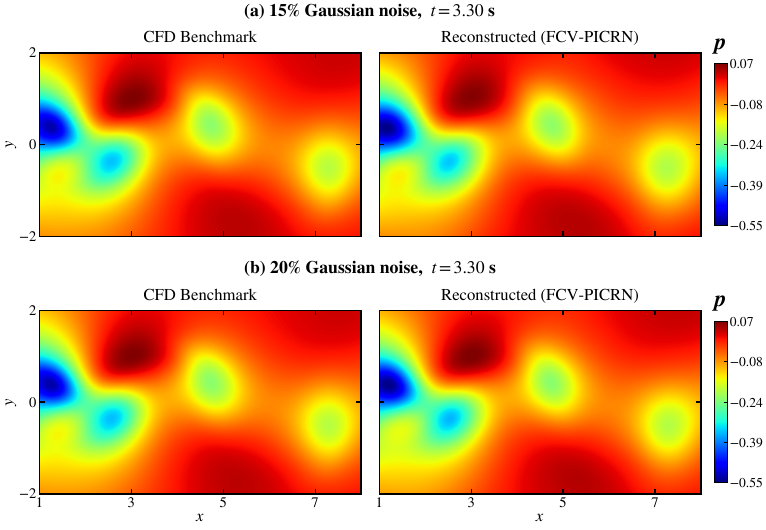}
\end{center}

\begin{quote}
\noindent\textbf{Figure \thefigure.}
Comparison of CFD Benchmark and FCV-PICRN reconstructed pressure for flow past a circular cylinder at $Re=100$ and $t=3.30$: (a) $15\%$ and (b) $20\%$ Gaussian noise. Pressure is unsupervised.
\end{quote}
\label{fig8}
\end{figure}

\begin{figure}[pos=!htbp]
\refstepcounter{figure}
\begin{center}
    \includegraphics[max width=\textwidth, max height=0.82\textheight]{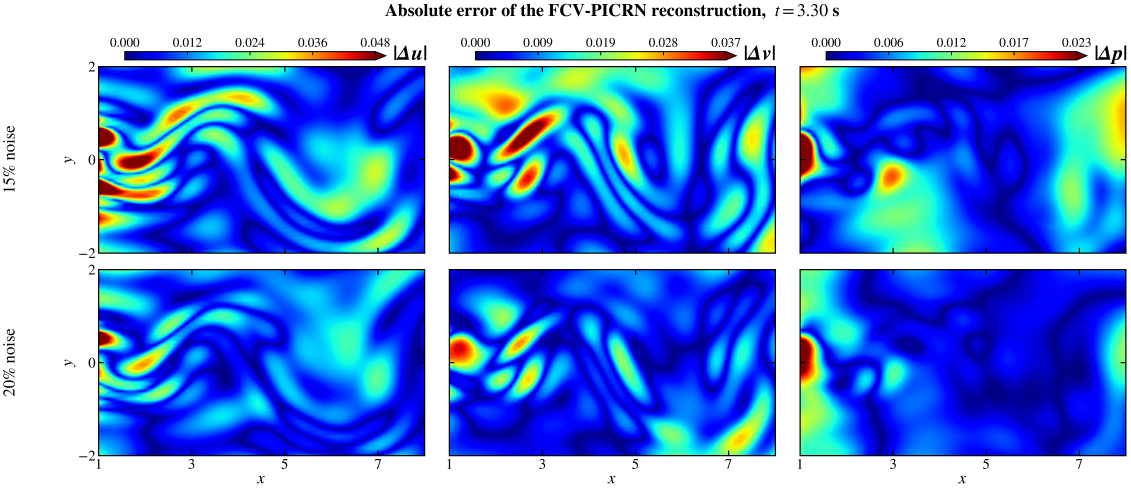}
\end{center}

\begin{quote}
\noindent\textbf{Figure \thefigure.}
Absolute errors of FCV-PICRN reconstructed $u$, $v$, and $p$ for flow past a circular cylinder at $Re=100$ and $t=3.30$: $15\%$ (top) and $20\%$ (bottom) Gaussian noise, respectively.
\end{quote}
\label{fig9}
\end{figure}

The third case assesses reconstruction from supervised velocity data corrupted by additive Gaussian noise (Model A, Equation~\eqref{eq:noise_gauss}) at $\eta=15\%$ and $20\%$. At $t=3.30$, Figure~\ref{fig7} shows strong contamination of the measured fields, particularly the lower-amplitude wake structures in $v$. Despite this, FCV-PICRN reconstructs smooth $u$ and $v$ fields that closely match the clean CFD benchmark, while the unsupervised pressure in Figures~\ref{fig8}a and \ref{fig8}b preserves the low-pressure vortex cores without visible noise contamination. The absolute errors in Figure~\ref{fig9} remain localized mainly near the inlet and shear layers, with maxima of approximately $0.05$, $0.04$, and $0.02$ for $u$, $v$, and $p$, respectively, at $15\%$ noise. Quantitatively, Table~\ref{Tab5} gives $R^{2}>0.997$, SSIM $>0.97$, and PSNR $\geq39.5$~dB for both noise levels. The PSNR values for $u$ and $v$ are $40.27$ and $39.55$~dB at $15\%$ noise and $42.74$ and $42.95$~dB at $20\%$ noise, respectively, demonstrating robust reconstruction under substantial Gaussian noise.

\begin{figure}[pos=!htbp]
\refstepcounter{figure}
\centering
\begin{minipage}[t]{0.49\textwidth}
\centering {\small\textbf{(a)}}\\[2pt]
\includegraphics[width=\linewidth]{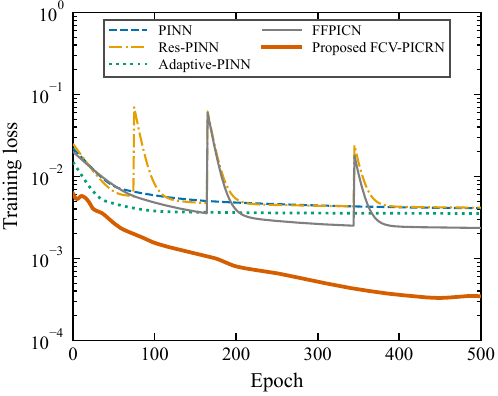}
\end{minipage}\hfill
\begin{minipage}[t]{0.49\textwidth}
\centering {\small\textbf{(b)}}\\[2pt]
\includegraphics[width=\linewidth]{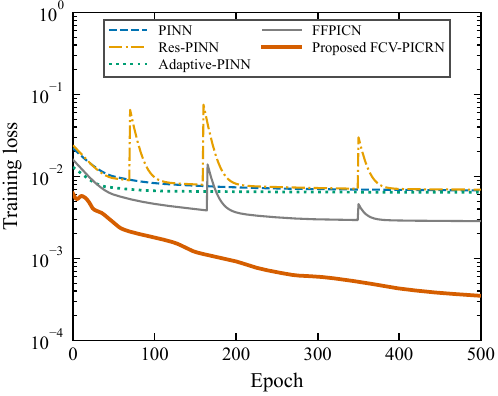}
\end{minipage}

\begin{quote}
\noindent\textbf{Figure \thefigure.}
Comparison of training-loss histories for PINN, Res-PINN, Adaptive-PINN, FFPICN, and the proposed FCV-PICRN for flow past a circular cylinder at $Re=100$: (a) $15\%$ and (b) $20\%$ Gaussian noise.
\end{quote}

\label{fig10}
\end{figure}

\begin{figure}[pos=!htbp]
\refstepcounter{figure}
\centering
\begin{minipage}[t]{0.49\textwidth}
\centering {\small\textbf{(a)}}\\[2pt]
\includegraphics[width=\linewidth]{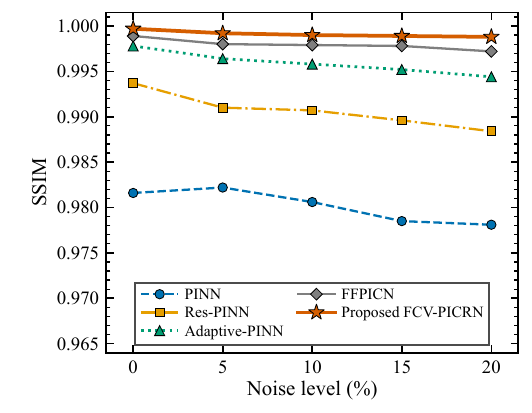}
\end{minipage}\hfill
\begin{minipage}[t]{0.49\textwidth}
\centering {\small\textbf{(b)}}\\[2pt]
\includegraphics[width=\linewidth]{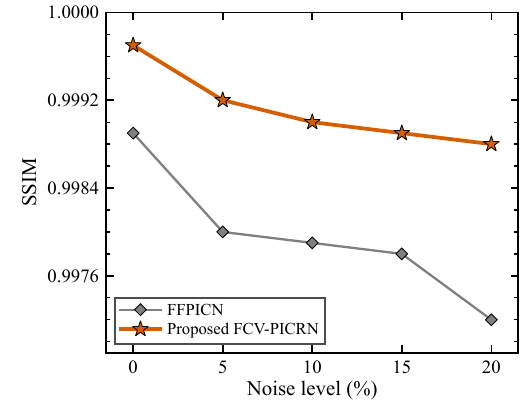}
\end{minipage}

\begin{quote}
\noindent\textbf{Figure \thefigure.}
Comparison of SSIM for flow past a circular cylinder at $Re=100$ under $0\%$--$20\%$ Gaussian noise: (a) PINN, Res-PINN, Adaptive-PINN, FFPICN, and the proposed FCV-PICRN; (b) FFPICN and the proposed FCV-PICRN.
\end{quote}

\label{fig11}
\end{figure}

The convergence and reconstruction accuracy of FCV-PICRN are compared with the conventional PINN, Res-PINN, Adaptive-PINN, and FFPICN~\cite{liu2025ffpicn} in Figures~\ref{fig10} and~\ref{fig11}. For both (a) $15\%$ and (b) $20\%$ Gaussian noise, Figure~\ref{fig10} shows that the FCV-PICRN loss decreases smoothly to approximately $4\times10^{-4}$ within 500 epochs, whereas the baseline losses remain between approximately $3\times10^{-3}$ and $8\times10^{-3}$, with pronounced spikes for Res-PINN and FFPICN. Since the loss functions differ in normalization, these histories indicate convergence behavior rather than direct reconstruction accuracy. Figure~\ref{fig11} therefore compares accuracy using SSIM. FCV-PICRN achieves the highest SSIM across the tested $0\%$--$20\%$ noise range (Figure~\ref{fig11}a), decreasing only from approximately $0.9997$ at $0\%$ to $0.9988$ at $20\%$, while the conventional PINN remains below $0.983$. The comparison with FFPICN in Figure~\ref{fig11}b further shows consistently higher SSIM for FCV-PICRN, with its value at $20\%$ noise ($\approx0.9988$) exceeding that of FFPICN at $5\%$ noise ($\approx0.9980$).

Table~\ref{TabNoise} provides the corresponding relative $L_2$ errors and $R^{2}$ values against FFPICN reported by Liu \textit{et al.}~\cite{liu2025ffpicn} for the same data set and noise levels. At $15\%$ noise, FCV-PICRN gives relative $L_2$ errors of $0.0015$, $0.0089$, and $0.0147$ for $u$, $v$, and $p$, respectively, representing reductions of $46\%$, $37\%$, and $40\%$ compared with FFPICN. At $20\%$ noise, the error in $v$ decreases by $61\%$ ($0.0075$ versus $0.0193$), while the pressure errors are comparable ($0.0244$ versus $0.0248$). FCV-PICRN also yields higher $R^{2}$ for all three fields at both noise levels, with values of at least $0.9993$. Together, these results demonstrate improved convergence and robust reconstruction under strong Gaussian noise.

\begin{table}[pos=htbp]
\centering
\caption{Comparison of FFPICN~\cite{liu2025ffpicn} and the proposed FCV-PICRN for flow past a circular cylinder at $Re=100$: relative $L_2$ error and $R^2$ under $15\%$ and $20\%$ Gaussian noise.}
\label{TabNoise}
\small\setlength{\tabcolsep}{4pt}
\begin{tabular}{llcccccc}
\hline
 & & \multicolumn{3}{c}{Relative $L_2$ error} & \multicolumn{3}{c}{$R^{2}$} \\
Method & Noise & $u$ & $v$ & $p$ & $u$ & $v$ & $p$ \\
\hline
FFPICN~\cite{liu2025ffpicn} & $15\%$ & 0.0028 & 0.0142 & 0.0247 & 0.9995 & 0.9993 & 0.9994 \\
 & $20\%$ & 0.0039 & 0.0193 & 0.0248 & 0.9994 & 0.9992 & 0.9992 \\
\textbf{FCV-PICRN} & $15\%$ & \textbf{0.0015} & \textbf{0.0089} & \textbf{0.0147} & \textbf{0.9998} & \textbf{0.9997} & \textbf{0.9995} \\
 & $20\%$ & \textbf{0.0028} & \textbf{0.0075} & \textbf{0.0244} & \textbf{0.9996} & \textbf{0.9997} & \textbf{0.9993} \\
\hline
\end{tabular}
\end{table}

\subsubsection{Cylinder-Wake Reconstruction at $Re=100$ from Critical-Wake Gappy Velocity Data with $5\%$ Gaussian and Speckle Noise}
\label{sec:critical_gaps}
The fourth case considers simultaneous data gaps and measurement noise in the critical near-wake region. Three gap geometries are examined: a large hole, $[2,4]\times[-1,1]$; a large truncation, $2\le x\le4$ over the full height; and a plus-shaped hole centered at $(3.5,0)$, with arms spanning $1.5\le x\le5.5$ and $-1.5\le y\le1.5$. No velocity data are supervised within these gaps. While Xu \textit{et al.}~\cite{xu2023practical} considered comparable holes and truncations using noise-free data, the present cases combine missing measurements with $5\%$ Gaussian noise; the large-truncation case is additionally tested with $5\%$ multiplicative speckle noise (Model B, Equation~\eqref{eq:noise_speckle}).

\begin{figure}[pos=!htbp]
\refstepcounter{figure}
\begin{center}
    \includegraphics[max width=\textwidth, max height=0.82\textheight]{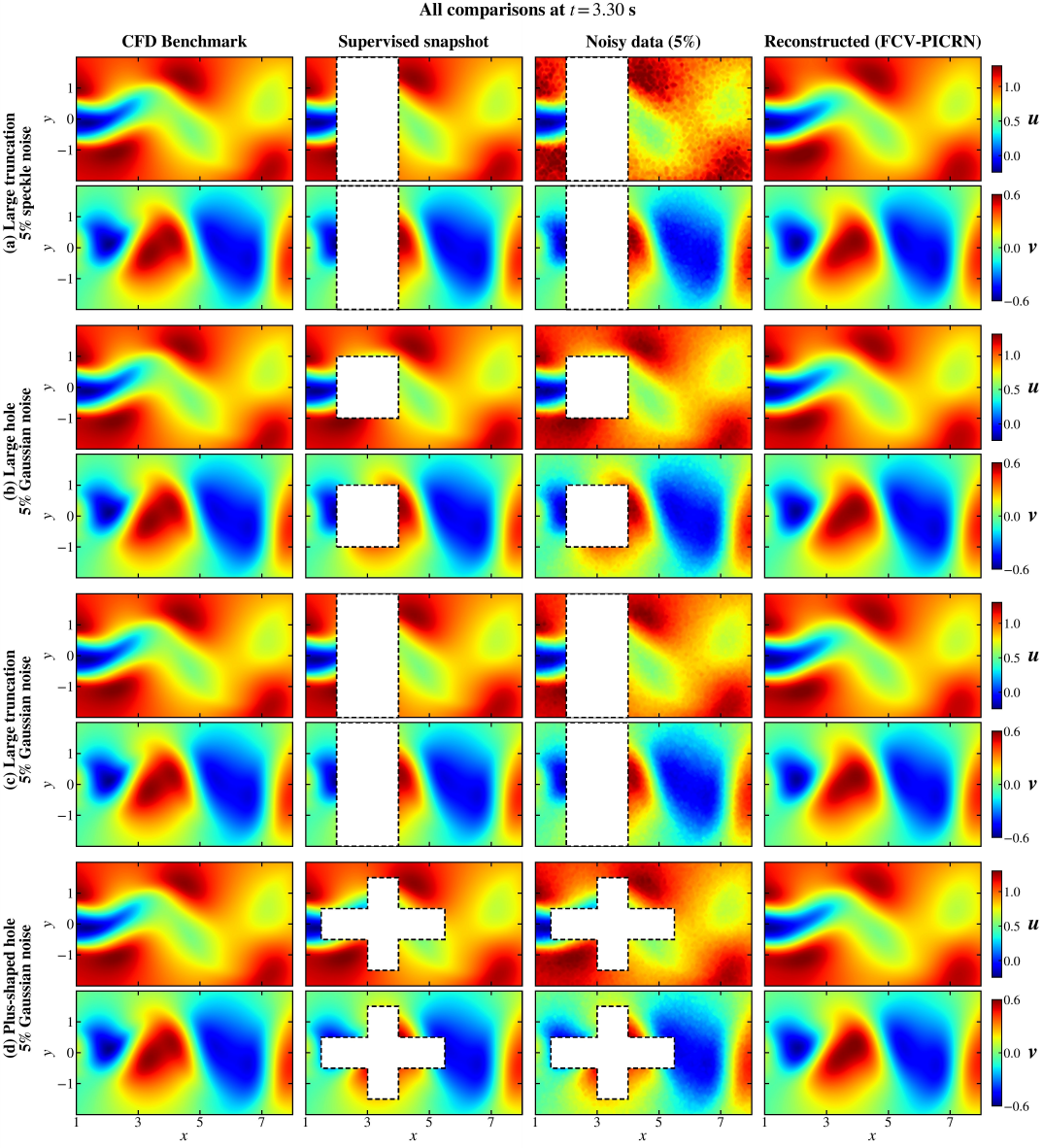}
\end{center}

\begin{quote}
\noindent\textbf{Figure \thefigure.}
CFD benchmark, supervised snapshot with gap, noisy data, and FCV-PICRN reconstruction of $u$ and $v$ for flow past a circular cylinder at $Re=100$ and $t=3.30$: (a) large truncation with $5\%$ speckle noise; (b) large hole, (c) large truncation, and (d) plus-shaped hole, each with $5\%$ Gaussian noise.
\end{quote}
\label{fig12}
\end{figure}

\begin{figure}[pos=!htbp]
\refstepcounter{figure}
\begin{center}
    \includegraphics[max width=\textwidth, max height=0.82\textheight]{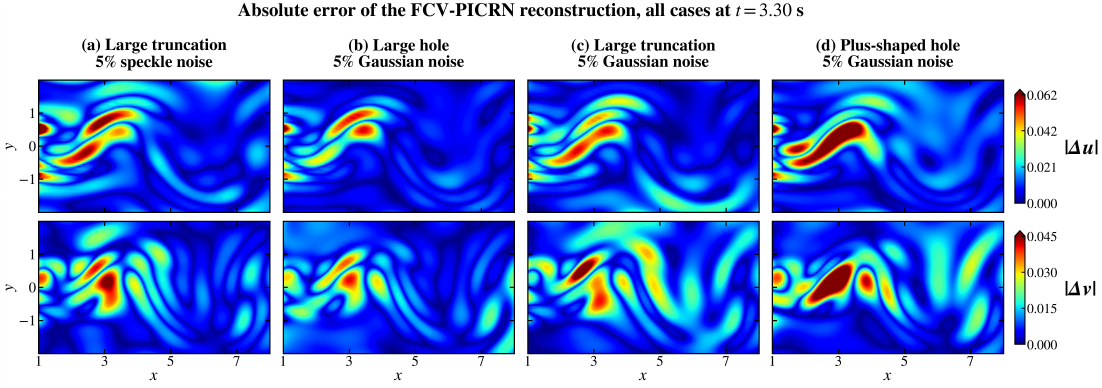}
\end{center}

\begin{quote}
\noindent\textbf{Figure \thefigure.}
Absolute errors of the reconstructed $u$ (top row) and $v$ (bottom row) for flow past a circular cylinder at $Re=100$ and $t=3.30$: (a) large truncation with $5\%$ speckle noise; (b) large hole, (c) large truncation, and (d) plus-shaped hole, each with $5\%$ Gaussian noise.
\end{quote}
\label{fig13}
\end{figure}

\begin{figure}[pos=!htbp]
\refstepcounter{figure}
\begin{center}
    \includegraphics[max width=\textwidth, max height=0.82\textheight]{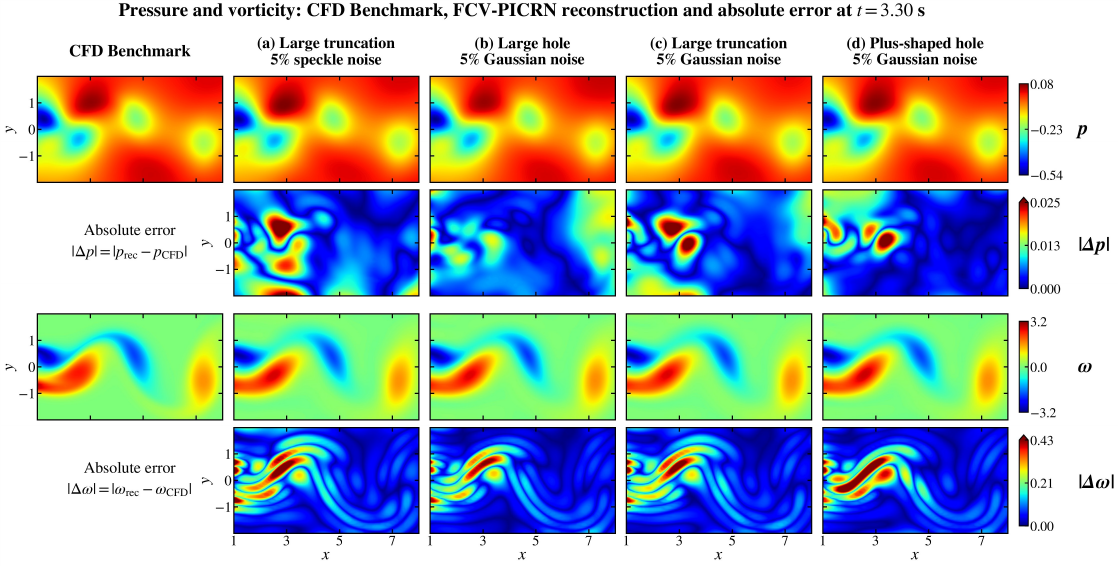}
\end{center}

\begin{quote}
\noindent\textbf{Figure \thefigure.}
Comparison of CFD Benchmark, FCV-PICRN reconstructed pressure and vorticity, and their absolute errors for flow past a circular cylinder at $Re=100$ and $t=3.30$: (a) large truncation with $5\%$ speckle noise; (b) large hole, (c) large truncation, and (d) plus-shaped hole, each with $5\%$ Gaussian noise. Neither pressure nor vorticity is supervised.
\end{quote}
\label{fig14}
\end{figure}

Figure~\ref{fig12} compares the supervised and reconstructed velocity fields at $t=3.30$. For all cases (a)--(d), FCV-PICRN reconstructs the missing wake structures and provides smooth continuity across the gaps. The absolute errors in Figure~\ref{fig13} are concentrated mainly along the shear layers within and downstream of the missing regions, with the plus-shaped gap (Figure~\ref{fig13}d) producing the largest velocity errors. The large-truncation results in Figures~\ref{fig13}a and \ref{fig13}c show comparable error magnitudes for speckle and Gaussian noise, although their spatial distributions differ. The unsupervised pressure and vorticity fields in Figure~\ref{fig14} also reproduce the principal wake structures, including the pressure minima associated with the vortex cores.

\begin{figure}[pos=!htbp]
\refstepcounter{figure}
\begin{center}
    \includegraphics[max width=\textwidth, max height=0.82\textheight]{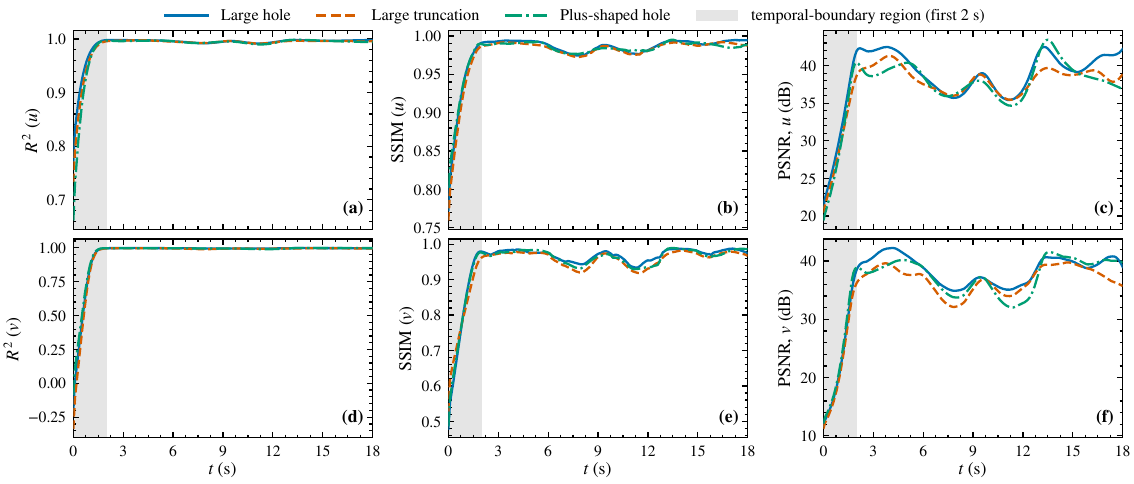}
\end{center}

\begin{quote}
\noindent\textbf{Figure \thefigure.}
Time histories of (a, d) $R^{2}$, (b, e) SSIM, and (c, f) PSNR of the reconstructed $u$ (a--c) and $v$ (d--f) for flow past a circular cylinder at $Re=100$ with three different gap configurations under $5\%$ Gaussian noise. The shaded band denotes the temporal-boundary region ($t<2$).
\end{quote}
\label{fig15}
\end{figure}

The time histories in Figure~\ref{fig15} show $R^{2}$ (a, d), SSIM (b, e), and PSNR (c, f) for the three gaps under $5\%$ Gaussian noise. For $t\ge2$, $R^{2}$ remains close to unity, SSIM exceeds approximately $0.97$ for $u$ and $0.92$ for $v$, and PSNR lies between approximately $32$ and $44$~dB; the lower accuracy for $t<2$ is associated with the temporal boundary. At $t=3.30$, Table~\ref{Tab5} gives $R^{2}>0.996$ for every gap. The PSNR values for $(u,v)$ are $(42.25,41.42)$~dB for the large hole, $(40.53,39.22)$~dB for the large truncation, and $(38.81,38.59)$~dB for the plus-shaped hole. For the large truncation with speckle noise, the corresponding values are $39.60$ and $40.65$~dB, respectively. These results demonstrate that FCV-PICRN maintains accurate reconstruction when missing critical-wake measurements are combined with Gaussian or multiplicative speckle noise.

\begin{table}[pos=htbp]
\centering
\caption{FCV-PICRN reconstruction metrics for noisy and gappy flow past a circular cylinder at $Re=100$ and $t=3.30$.}
\label{Tab5}
\small
\begin{tabular}{llcccccc}
\hline
 & & \multicolumn{2}{c}{$R^{2}$} & \multicolumn{2}{c}{SSIM} & \multicolumn{2}{c}{PSNR (dB)} \\
Gap & Noise & $u$ & $v$ & $u$ & $v$ & $u$ & $v$ \\
\hline
None & $15\%$ Gaussian & 0.9973 & 0.9979 & 0.9905 & 0.9706 & 40.27 & 39.55 \\
None & $20\%$ Gaussian & 0.9984 & 0.9990 & 0.9955 & 0.9876 & 42.74 & 42.95 \\
Large hole & $5\%$ Gaussian & 0.9983 & 0.9986 & 0.9939 & 0.9822 & 42.25 & 41.42 \\
Large truncation & $5\%$ Gaussian & 0.9974 & 0.9977 & 0.9916 & 0.9747 & 40.53 & 39.22 \\
Large truncation & $5\%$ speckle & 0.9968 & 0.9983 & 0.9893 & 0.9809 & 39.60 & 40.65 \\
Plus-shaped hole & $5\%$ Gaussian & 0.9962 & 0.9973 & 0.9903 & 0.9765 & 38.81 & 38.59 \\
\hline
\end{tabular}
\end{table}

\subsubsection{Cylinder-Wake Reconstruction at $Re=100$ from Plus-Shaped Gappy Velocity Data under $5\%$, $10\%$, and $15\%$ Gaussian Noise}
\label{sec:plus_noise}

\begin{figure}[pos=!htbp]
\refstepcounter{figure}
\begin{center}
    \includegraphics[max width=\textwidth, max height=0.55\textheight]{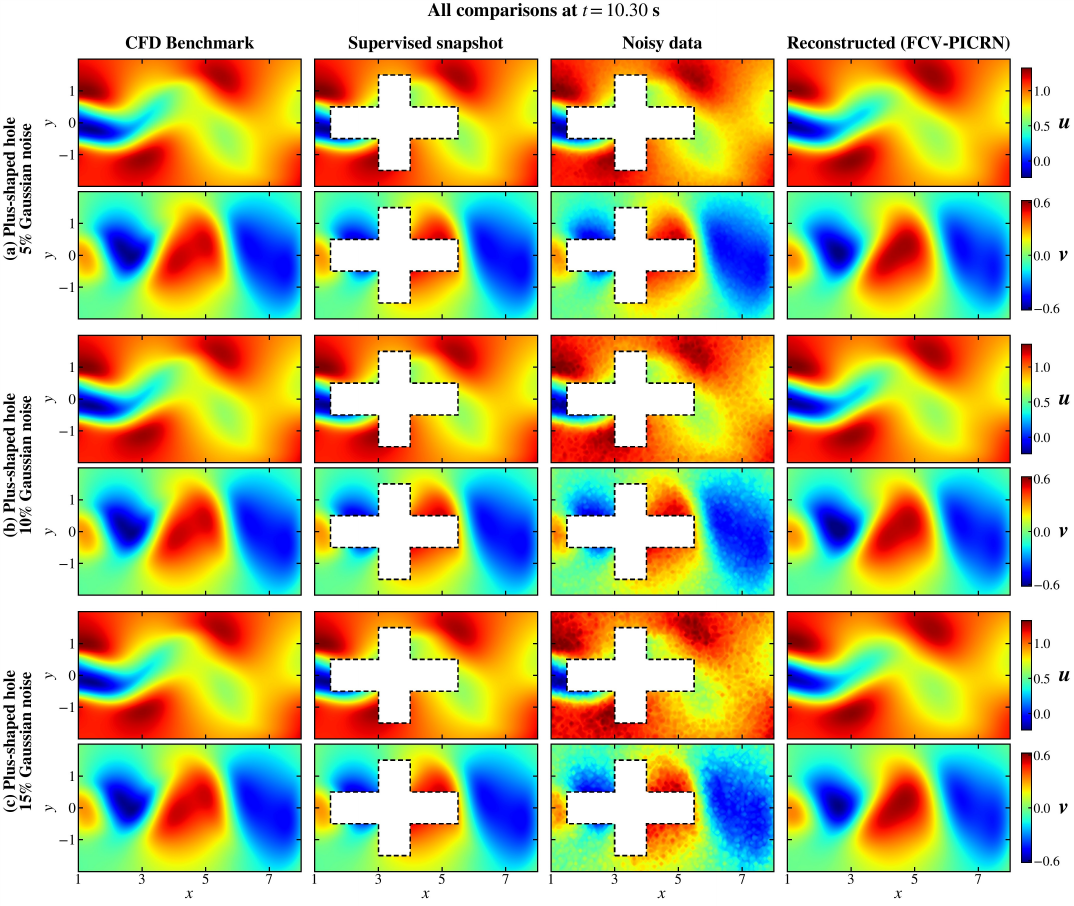}
\end{center}

\begin{quote}
\noindent\textbf{Figure \thefigure.}
CFD Benchmark, supervised snapshot, noisy data, and FCV-PICRN reconstruction of $u$ and $v$ for flow past a circular cylinder at $Re=100$ and $t=10.30$ with a plus-shaped hole under (a) $5\%$, (b) $10\%$, and (c) $15\%$ Gaussian noise.
\end{quote}
\label{fig16}
\end{figure}

\begin{figure}[pos=!htbp]
\refstepcounter{figure}
\begin{center}
    \includegraphics[max width=\textwidth, max height=0.82\textheight]{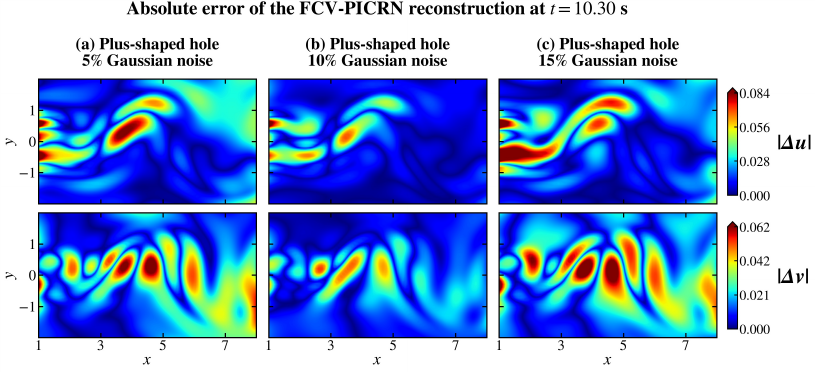}
\end{center}

\begin{quote}
\noindent\textbf{Figure \thefigure.}
Absolute errors of the reconstructed $u$ (top row) and $v$ (bottom row) for flow past a circular cylinder at $Re=100$ and $t=10.30$ with a plus-shaped hole under $5\%$, $10\%$, and $15\%$ Gaussian noise, respectively.
\end{quote}

\label{fig17}
\end{figure}

\begin{figure}[pos=!htbp]
\refstepcounter{figure}
\begin{center}
    \includegraphics[max width=\textwidth, max height=0.82\textheight]{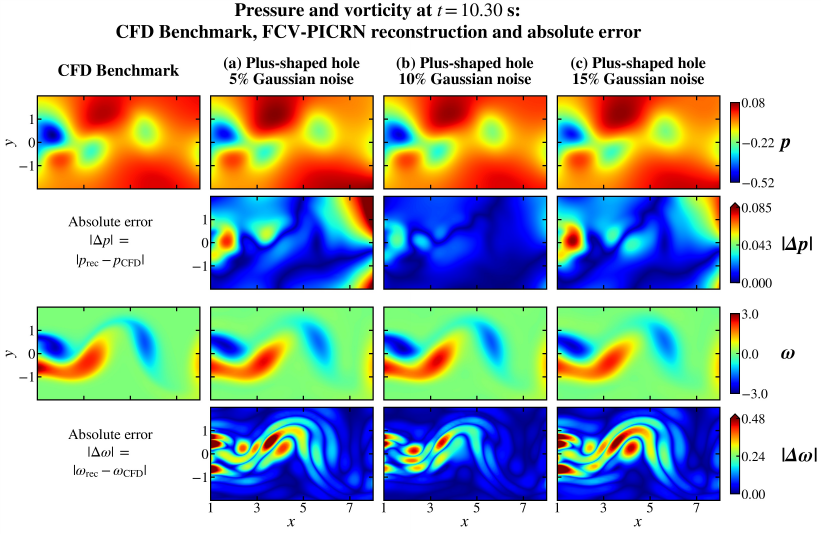}
\end{center}

\begin{quote}
\noindent\textbf{Figure \thefigure.}
Comparison of CFD benchmark, FCV-PICRN reconstructed pressure and vorticity, and their absolute errors for flow past a circular cylinder at $Re=100$ and $t=10.30$ with a plus-shaped hole under $5\%$, $10\%$, and $15\%$ Gaussian noise, respectively.
\end{quote}

\label{fig18}
\end{figure}

The final case evaluates the effect of Gaussian noise at $\eta=5\%$, $10\%$, and $15\%$ for the plus-shaped gap. Figure~\ref{fig16} compares the supervised and reconstructed velocity fields at $t=10.30$. FCV-PICRN reconstructs the vortex street within and around the gap at all three noise levels, including the strongly contaminated $15\%$ case (Figure~\ref{fig16}c). The absolute errors in Figures~\ref{fig17}a--c increase moderately with noise and remain concentrated mainly within the vortex-formation region, with maxima of approximately $0.08$ for $u$ and $0.06$ for $v$. The corresponding unsupervised pressure and vorticity fields in Figures~\ref{fig18}a--c also preserve the principal wake structures across all noise levels.

\begin{figure}[pos=!htbp]
\refstepcounter{figure}
\begin{center}
    \includegraphics[max width=\textwidth, max height=0.82\textheight]{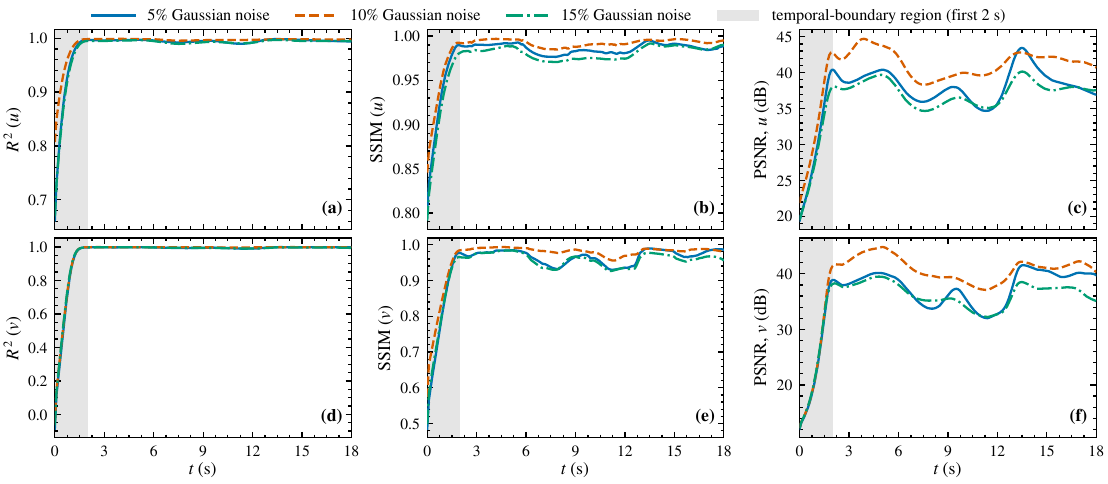}
\end{center}

\begin{quote}
\noindent\textbf{Figure \thefigure.}
Time histories of (a, d) $R^{2}$, (b, e) SSIM, and (c, f) PSNR of the reconstructed $u$ (a--c) and $v$ (d--f) for flow past a circular cylinder at $Re=100$ with a plus-shaped hole under $5\%$, $10\%$, and $15\%$ Gaussian noise. The shaded band marks the temporal-boundary region ($t<2$).
\end{quote}
\label{fig19}
\end{figure}

The time histories in Figure~\ref{fig19} indicate stable reconstruction for all three cases, with $R^{2}$ close to unity and PSNR between approximately $32$ and $45$~dB for $t\ge2$. At $t=10.30$, Table~\ref{TabPlus} gives $R^{2}>0.99$ for all noise levels. As $\eta$ increases from $5\%$ to $15\%$, SSIM decreases from $0.9825$ to $0.9749$ for $u$ and from $0.9593$ to $0.9512$ for $v$, while the PSNR of $v$ decreases from $34.53$ to $33.47$~dB. For $u$, the PSNR changes from $36.19$~dB at $5\%$ to $36.89$~dB at $10\%$, showing only a small non-monotonic variation. Overall, increasing the Gaussian noise from $5\%$ to $15\%$ causes only a modest reduction in reconstruction accuracy for the plus-shaped gap.

\begin{table}[pos=htbp]
\centering
\caption{FCV-PICRN reconstruction metrics for flow past a circular cylinder at $Re=100$ with a plus-shaped hole under $5\%$, $10\%$, and $15\%$ Gaussian noise at $t=10.30$.}
\label{TabPlus}
\small
\begin{tabular}{lcccccc}
\hline
 & \multicolumn{2}{c}{$R^{2}$} & \multicolumn{2}{c}{SSIM} & \multicolumn{2}{c}{PSNR (dB)} \\
Noise & $u$ & $v$ & $u$ & $v$ & $u$ & $v$ \\
\hline
$5\%$ Gaussian & 0.9927 & 0.9936 & 0.9825 & 0.9593 & 36.19 & 34.53 \\
$10\%$ Gaussian & 0.9938 & 0.9936 & 0.9793 & 0.9583 & 36.89 & 34.49 \\
$15\%$ Gaussian & 0.9923 & 0.9918 & 0.9749 & 0.9512 & 35.94 & 33.47 \\
\hline
\end{tabular}
\end{table}

\subsection{Two-Dimensional Turbulence Reconstruction from Sparse Velocity Data with $5\%$ Gaussian Noise}
\label{sec:turbulence}

The third test case considers statistically stationary two-dimensional turbulence driven by steady Kolmogorov forcing and damped by linear bottom friction, representative of thin-layer laboratory flows~\cite{xia2009}. This configuration was introduced by Parfenyev \textit{et al.}~\cite{parfenyev2024} for PINN-based turbulent-flow reconstruction. The corresponding openly available reference data set~\cite{parfenyev2024data} is used here without modification.

\subsubsection{Problem Statement and Data}
\label{sec:turb_problem}

The flow is governed by the forced incompressible Navier--Stokes equations with linear friction,
\begin{equation}
  \partial_t\mathbf{u}+(\mathbf{u}\cdot\nabla)\mathbf{u}
  =-\nabla p-\alpha\,\mathbf{u}+\nu\nabla^{2}\mathbf{u}
  +\mathbf{f}(\mathbf{x}),\qquad
  \nabla\cdot\mathbf{u}=0,
  \label{eq:kolmogorov}
\end{equation}
where $\mathbf{u}=(u,v)$, $\alpha$ is the bottom-friction coefficient, $\nu$ is the kinematic viscosity, and $\mathbf{f}=(f_0\sin k_f y,\,0)$ is the Kolmogorov force with $f_0=10$ and $k_f=5$. For the root-mean-square velocity $U\approx4.1$ and window size $L_0=\pi$, the corresponding Reynolds numbers are $Re=UL_0/\nu\approx1.3\times10^{3}$ and $Rh=U/(\alpha L_0)\approx13$~\cite{mishra2015,parfenyev2024}. The reference DNS of Ref.~\cite{parfenyev2024} was generated using the pseudo-spectral solver GeophysicalFlows.jl~\cite{constantinou2021} in a doubly periodic $2\pi\times2\pi$ domain with $512^{2}$ grid points, $\nu=0.01$, $\alpha=0.1$, and two-thirds dealiasing. After the initial transient, 200 snapshots were recorded with $\Delta t=0.02$ over $T=4$, corresponding to approximately five large-eddy turnover times, $L_0/U\approx0.77$.

\begin{figure}[pos=!htbp]
\refstepcounter{figure}
\begin{center}
    \includegraphics[max width=\textwidth, max height=0.82\textheight]{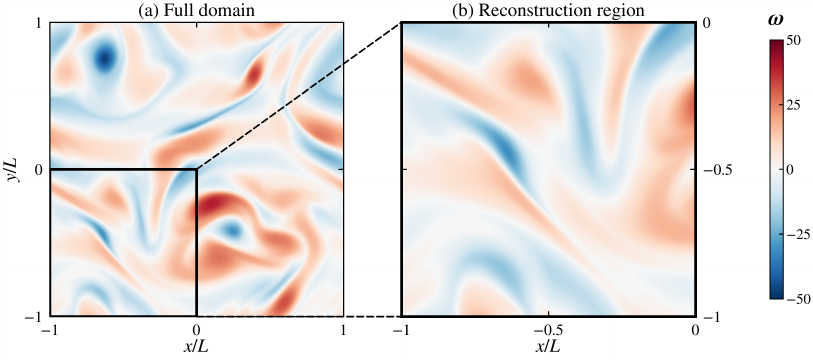}
\end{center}

\begin{quote}
\noindent\textbf{Figure \thefigure.}
Vorticity of the forced two-dimensional turbulence: (a) the full DNS domain and (b) the observation window $\Omega=[-\pi,0]^{2}$ used for reconstruction. The axes are scaled by the window size $L_0=\pi$ (labelled $L$ in the panels).
\end{quote}

\label{fig20}
\end{figure}

The reconstruction uses the observation window $\Omega=[-\pi,0]^{2}$ with $256\times256$ grid points (Figure~\ref{fig20}), without prescribing boundary conditions on $\partial\Omega$. Unlike the cylinder-wake cases, no spatial gap is introduced. Instead, $N_d=2\times10^{5}$ randomly sampled space--time points, approximately $1.5\%$ of the grid, provide velocity supervision corrupted by $5\%$ additive Gaussian noise (Model A, Equation~\eqref{eq:noise_gauss} with $\eta=0.05$). The pressure $p$, force $\mathbf{f}$, friction coefficient $\alpha$, and viscosity $\nu$ are treated as unknown. Following Ref.~\cite{parfenyev2024}, positive parameter estimates are imposed as $\hat{\alpha}=e^{\zeta_1}$ and $\hat{\nu}=e^{\zeta_2}$, where $\zeta_1$ and $\zeta_2$ are trainable scalars. The unknown force is represented by a time-independent real network $G_{\theta_f}:(x,y)\mapsto(\hat{f}_x,\hat{f}_y)$ with parameters $\theta_f$, and its divergence is penalized to enforce a solenoidal inferred force (Equation~\eqref{eq:turb_residuals}). Because the open window admits the gauge transformations $p\to p+h(\mathbf{x})+c(t)$ and $\mathbf{f}\to\mathbf{f}+\nabla h$, with $\nabla^{2}h=0$~\cite{parfenyev2024}, pressure is evaluated after removing its spatial mean at each snapshot.

\subsubsection{Loss Function and Network Configurations}
\label{sec:turb_setup}
For the turbulent-flow case, Equations~\eqref{eq:e1}--\eqref{eq:e3} are augmented by linear friction, forcing, and force-divergence terms and nondimensionalized using $U$ and $L_0$ following Ref.~\cite{parfenyev2024}:
\begin{equation}
\begin{aligned}
  e_1&=\left[\hat{u}_t+u\hat{u}_x+v\hat{u}_y+\hat{p}_x+\hat{\alpha}\hat{u}-\hat{\nu}(\hat{u}_{xx}+\hat{u}_{yy})-\hat{f}_x\right]\frac{L_0}{U^{2}}, &
  e_3&=(\hat{u}_x+\hat{v}_y)\frac{L_0}{U},\\
  e_2&=\left[\hat{v}_t+u\hat{v}_x+v\hat{v}_y+\hat{p}_y+\hat{\alpha}\hat{v}-\hat{\nu}(\hat{v}_{xx}+\hat{v}_{yy})-\hat{f}_y\right]\frac{L_0}{U^{2}}, &
  e_4&=(\partial_x\hat{f}_x+\partial_y\hat{f}_y)\frac{L_0^{2}}{U^{2}}.
\end{aligned}
\label{eq:turb_residuals}
\end{equation}
The corresponding physics loss is
$\mathcal{L}_{\mathrm{physics}}=N_c^{-1}\sum_j\left(|e_{1,j}|^{2}+|e_{2,j}|^{2}+|e_{3,j}|^{2}+e_{4,j}^{2}\right)$,
and the total objective is
$\mathcal{L}=\mathcal{L}_{\mathrm{data}}+\lambda\mathcal{L}_{\mathrm{physics}}$,
where $\mathcal{L}_{\mathrm{data}}$ is the variance-normalized misfit defined in Equation~\eqref{eq:loss_data} and $\lambda=0.02$~\cite{parfenyev2024}. Collocation points are selected by Latin hypercube sampling from a pool of $10^{6}$ points over $\Omega\times[0,T]$. To reduce computational cost, the derivatives in Equation~\eqref{eq:turb_residuals} are evaluated using second-order central differences of the network function in normalized coordinates, with spatial and temporal steps of $2^{-10}$ and $2^{-12}$, respectively. The resulting derivatives agree with automatic differentiation within a relative $L_2$ error of approximately $10^{-3}$, substantially below the imposed noise level.

\begin{inlinetable}
\caption{Hyperparameters of FCV-PICRN and the PINN baseline~\cite{parfenyev2024} for the forced two-dimensional turbulence.}
\label{Tab6}
\small
\begin{tabular}{lcc}
\hline
Parameter & FCV-PICRN & PINN \\
\hline
Fourier order $K$ & 6 & -- \\
Hidden width $W$ / blocks (layers) $L$ & 256 / 8 & 108 / 6 \\
Activation / initialization & CGELU / polar & $\tanh$ / Xavier \\
Trainable parameters & 71,686 & 71,391 \\
\hline
\multicolumn{3}{l}{\textit{Shared settings}} \\
Force network $G_{\theta_f}$ & \multicolumn{2}{c}{$3\times32$, $\tanh$} \\
Data points $N_d$ & \multicolumn{2}{c}{$2\times10^{5}$ ($5\%$ Gaussian noise)} \\
Collocation points & \multicolumn{2}{c}{$10^{6}$ (LHS)} \\
Batch size & \multicolumn{2}{c}{4096} \\
Physics weight $\lambda$ & \multicolumn{2}{c}{0.02} \\
Phase I: Adam~\cite{kingma2014adam} & \multicolumn{2}{c}{$3\times10^{4}$ iterations, lr $2\times10^{-3}\rightarrow4\times10^{-5}$} \\
Phase II: SSBroyden~\cite{urban2025} & \multicolumn{2}{c}{600 iterations} \\
\hline
\end{tabular}
\end{inlinetable}

FCV-PICRN is compared with the residual $\tanh$ PINN of Ref.~\cite{parfenyev2024}, with the baseline width selected to give comparable trainable parameter counts (71,391 versus 71,686). Both models use the same force network, training data, collocation pool, $\lambda$, optimizer settings, and stopping criteria; the configurations are summarized in Table~\ref{Tab6}. For FCV-PICRN, the hidden width $W=256$ is reshaped into 32 channels of length 8, extending Equation~\eqref{eq:cconv} to multiple channels. In both models, the outputs $\ReP(\hat{u},\hat{v},\hat{p})$ are scaled by $(U,U,U^{2})$, while the FCV-PICRN output-layer weights are initialized at one-tenth of their standard magnitude. The force network comprises three $\tanh$ layers with 32 neurons each (2,274 parameters), with $\zeta_1^{(0)}=0$ and $\zeta_2^{(0)}=-3$. Adam employs learning-rate warm-up followed by cosine decay, a tenfold larger learning rate for $\zeta_{1,2}$, and unit-norm gradient clipping. SSBroyden uses curvature memory $m=30$, a search-direction bound $\|\mathbf{d}\|\le1$, and a strong-Wolfe line search. All computations are performed in single precision on two NVIDIA T4 GPUs.

\begin{figure}[pos=!htbp]
\refstepcounter{figure}
\begin{center}
    \includegraphics[max width=0.92\textwidth, max height=0.82\textheight]{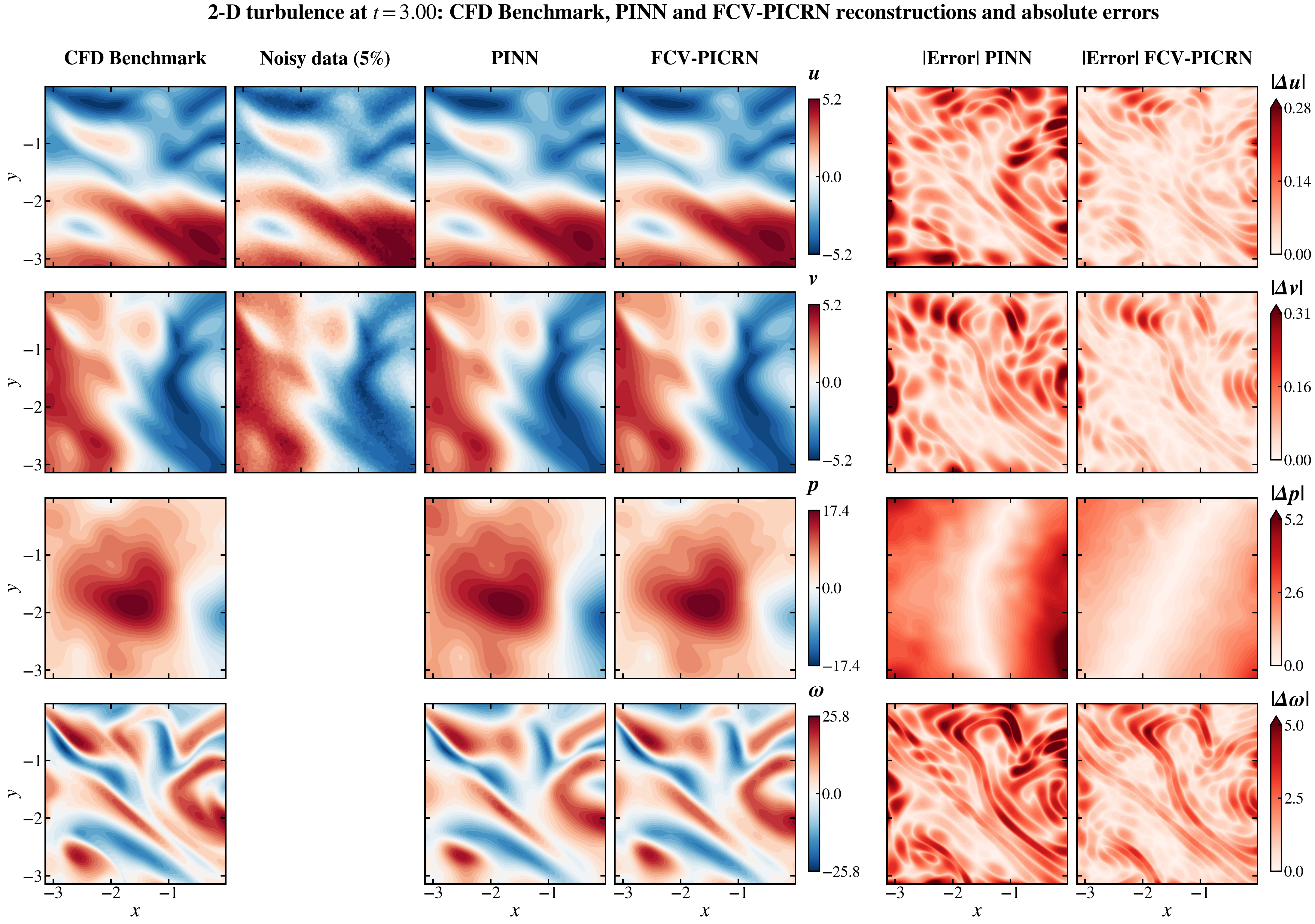}
\end{center}

\begin{quote}
\noindent\textbf{Figure \thefigure.}
Forced two-dimensional turbulence at $t=3.00$: CFD Benchmark(DNS), noisy data ($5\%$ Gaussian), PINN and FCV-PICRN reconstructions of $u$, $v$, $p$, and $\omega$, and the absolute errors of both models.
\end{quote}

\label{fig21}
\end{figure}

\subsubsection{Results}
\label{sec:turb_results}
Figure~\ref{fig21} compares the reconstructed fields with the DNS at $t=3.00$. Both models suppress the measurement noise and reproduce the large-scale velocity structures; however, FCV-PICRN yields smaller absolute errors, concentrated mainly along thin shear layers, whereas the PINN errors extend across the observation window, with $|\Delta u|$ and $|\Delta v|$ frequently approaching $0.3$. The difference is more evident for the unsupervised fields: FCV-PICRN reduces the pressure error, particularly near the lateral boundaries, and more accurately reconstructs the vorticity and its thin filamentary structures.

\begin{figure}[pos=!htbp]
\refstepcounter{figure}
\begin{center}
    \includegraphics[max width=\textwidth, max height=0.82\textheight]{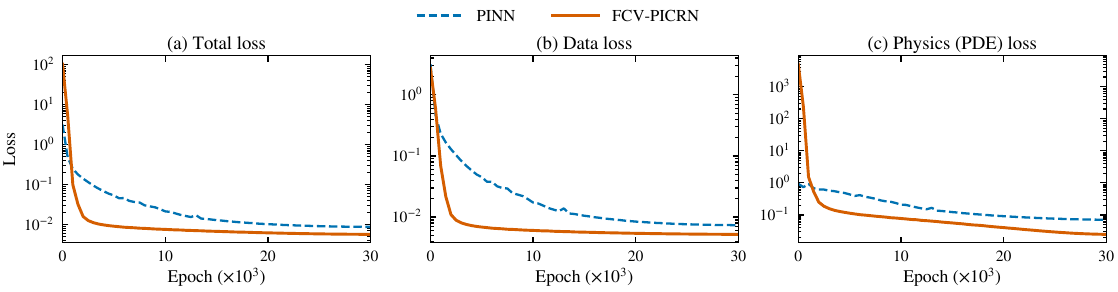}
\end{center}

\begin{quote}
\noindent\textbf{Figure \thefigure.}
Training-loss histories of the PINN and FCV-PICRN for the forced turbulence: (a) total loss, (b) data loss, and (c) physics loss.
\end{quote}

\label{fig22}
\end{figure}

\begin{figure}[pos=!htbp]
\refstepcounter{figure}
\begin{center}
    \includegraphics[max width=\textwidth, max height=0.82\textheight]{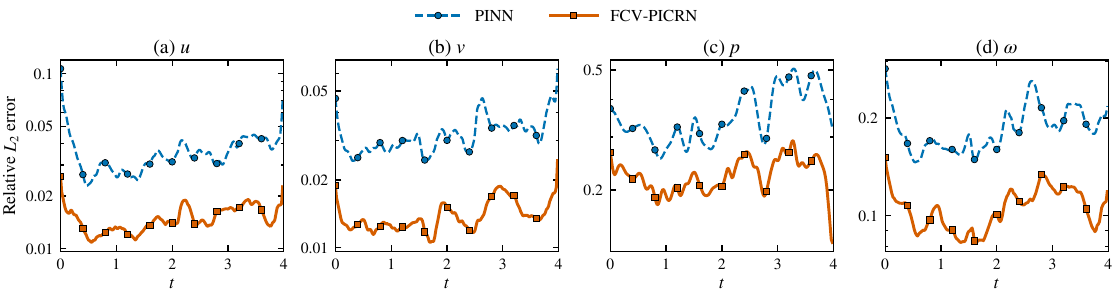}
\end{center}

\begin{quote}
\noindent\textbf{Figure \thefigure.}
Relative $L_2$ errors (Equation~\eqref{eq:metrics1}) of the PINN and FCV-PICRN reconstructions over time for (a) $u$, (b) $v$, (c) $p$, and (d) $\omega$.
\end{quote}

\label{fig23}
\end{figure}

The convergence histories in Figure~\ref{fig22} show that the total and data losses of FCV-PICRN (Figures~\ref{fig22}a and \ref{fig22}b) decrease by more than two orders of magnitude within the first $3\times10^{3}$ epochs, while the PINN converges more gradually. FCV-PICRN also reaches lower final values for all three losses, with a physics loss (Figure~\ref{fig22}c) approximately $3\times$ smaller than that of the PINN. Consistent with these trends, Figure~\ref{fig23} shows lower relative $L_2$ errors throughout the time interval. The FCV-PICRN errors for $u$ and $v$ (Figures~\ref{fig23}a and \ref{fig23}b) remain approximately $1\%$--$2\%$, compared with $2.5\%$--$5\%$ for the PINN. The pressure error (Figure~\ref{fig23}c) decreases from approximately $0.3$--$0.5$ to $0.18$--$0.28$, while the vorticity error (Figure~\ref{fig23}d) decreases from approximately $0.16$--$0.24$ to $0.08$--$0.14$. The larger errors near $t=0$ and $t=T$ are consistent with the temporal-boundary effects observed for the cylinder wake. Since the models have comparable parameter counts and use identical training conditions, these results indicate that the Fourier embedding and complex convolutional-residual representation improve the reconstruction of broadband turbulent flow with unknown forcing and dissipation.

\FloatBarrier
\section{Conclusion}
\label{sec:conclusion}

This work developed a Fourier-feature-embedded complex-valued physics-informed convolutional residual network (FCV-PICRN) for reconstructing incompressible flow fields from sparse, gappy, and noisy velocity measurements, particularly when observations are completely unavailable in dynamically important regions. The framework combines Fourier features to represent oscillatory and multiscale structures, complex-valued latent variables for coupled amplitude--phase learning, convolutional residual blocks for compact feature processing, and governing-equation constraints to guide the reconstruction in both observed and unobserved regions; training is performed using Adam followed by self-scaled quasi-Newton refinement. The Taylor--Green vortex at $Re=100$ first verifies the method under known physical constraints, while the cylinder wake at $Re=100$ provides a substantially more challenging reconstruction setting involving extremely sparse measurements, large-hole, truncation, and plus-shaped gaps within the vortex-formation region, together with additive Gaussian and multiplicative speckle noise. FCV-PICRN reconstructs velocity, pressure, and vorticity within these unsupervised regions, remains robust under Gaussian noise levels up to $20\%$, and provides improved accuracy over the reported FFPICN results~\cite{liu2025ffpicn} under comparable sparse-data settings. Its performance on forced two-dimensional turbulence with $5\%$ Gaussian noise further demonstrates its ability to reconstruct broadband flow dynamics, with lower reconstruction errors than a conventional PINN of comparable size. These results show that integrating frequency-aware encoding, complex-valued feature learning, convolutional residual processing, and physical constraints provides an effective framework for simultaneously addressing sparsity, structured data gaps, and measurement noise across laminar, wake, and turbulent flows. Future work will focus on hard boundary and periodic constraints, statistical assessment over multiple noise realizations and independent training runs, extension to three-dimensional and higher-Reynolds-number turbulent flows, and validation using experimental PIV/PTV measurements with realistic occlusions and measurement uncertainty.

\section*{Declaration of Generative AI and AI-assisted Technologies in the Writing Process}
During the preparation of this manuscript, the authors used ChatGPT Plus solely for language polishing and grammatical correction. After using this tool, the authors carefully reviewed, revised, and validated the content. The authors take full responsibility for the scientific integrity and content of the published article.

\section*{Author Declaration}
The authors have no conflicts of interest to disclose.

\section*{Data Availability}
The data that support the findings of this study are available from the corresponding author upon reasonable request.

\end{document}